\documentclass[backref, breaklinks, onecolumn, colorlinks]{aastex63}   
\hypersetup{linkcolor=blue,citecolor=blue,filecolor=cyan,urlcolor=magenta}

\usepackage{enumitem}
\usepackage{mathtools}
\usepackage{graphicx}
\usepackage{amssymb}
\usepackage{amsmath}
\usepackage{ulem}
\usepackage{epstopdf}
\usepackage{color}
\usepackage{graphicx}
\usepackage{xcolor}

\def\lambar{\lambda\llap {--}}

{\catcode`\@=11
\gdef\SchlangeUnter#1#2{\lower2pt\vbox{\baselineskip 0pt\lineskip0pt
\ialign{$\m@th#1\hfil##\hfil$\crcr#2\crcr\sim\crcr}}}}
\def\gtrsim{\mathrel{\mathpalette\SchlangeUnter>}}
\def\lesssim{\mathrel{\mathpalette\SchlangeUnter<}}

\begin{document}
\tighten

\title{The Fast Radio Burst Luminosity Function and Death Line in the Low-Twist Magnetar Model}

\author[0000-0002-9249-0515]{Zorawar Wadiasingh}
\affil{Astrophysics Science Division, NASA Goddard Space Flight Center, Greenbelt, Maryland, 20771, USA}
\affil{Universities Space Research Association (USRA) Columbia, Maryland 21046, USA}
\affil{Centre for Space Research, North-West University, Potchefstroom, South Africa }

\author{Paz Beniamini}
\affil{Department of Physics, The George Washington University, 725 21st St. NW, Washington, DC 20052, USA}
\affil{TAPIR, Mailcode 350-17, California Institute of Technology, Pasadena, CA 91125, USA}

\author{Andrey Timokhin}
\affil{Janusz Gil Institute of Astronomy, University of Zielona G\'ora, ul. Szafrana 2, 65-516 Zielona G\'ora, Poland}
\affil{Astrophysics Science Division, NASA Goddard Space Flight Center, Greenbelt, Maryland, 20771, USA}

\author{Matthew G. Baring}
\affil{Department of Physics and Astronomy - MS 108, Rice University,
6100 Main Street, Houston, Texas 77251-1892, USA}

\author{Alexander J. van der Horst}
\affil{Department of Physics, The George Washington University, 725 21st St. NW, Washington, DC 20052, USA}
\affil{Astronomy, Physics and Statistics Institute of Sciences (APSIS), The George Washington University, Washington, DC 20052, USA}

\author{Alice K. Harding}
\affil{Astrophysics Science Division, NASA Goddard Space Flight Center, Greenbelt, Maryland, 20771, USA}

\author{Demosthenes Kazanas}
\affil{Astrophysics Science Division, NASA Goddard Space Flight Center, Greenbelt, Maryland, 20771, USA}

\begin{abstract}
We explore the burst energy distribution of fast radio bursts (FRBs) in the low-twist magnetar model of \cite{2019ApJ...879....4W}. Motivated by the power-law fluence distributions of FRB~121102, we propose an elementary model for the FRB luminosity function of individual repeaters with an inversion protocol which directly relates the power-law distribution index of magnetar short burst fluences to that for FRBs. The protocol indicates the FRB energy scales virtually linearly with crust/field dislocation amplitude, if magnetar short bursts prevail in the magnetoelastic regime. Charge starvation in the magnetosphere during bursts (required in WT19) for individual repeaters implies the predicted burst fluence distribution is narrow, $\lesssim 3$ decades for yielding strains and oscillation frequencies feasible in magnetar crusts. Requiring magnetic confinement and charge starvation, we obtain a death line for FRBs which segregates magnetars from the normal pulsar population, suggesting only the former will host recurrent FRBs. We convolve the burst energy distribution for individual magnetars to define the distribution of luminosities in evolved magnetar populations. The broken power-law luminosity function's low energy character depends on the population model, while the high energy index traces that of individual repeaters. Independent of the evolved population, the broken power-law isotropic-equivalent energy/luminosity function peaks at $\sim10^{37}-10^{40}$~erg with a low-energy cutoff at $\sim 10^{37}$~erg.  Lastly, we consider the local fluence distribution of FRBs, and find that it can constrain the subset of FRB-producing magnetar progenitors. Our model suggests that improvements in sensitivity may reveal flattening of the global FRB fluence distribution and saturation in FRB rates.
\end{abstract} 

\section{Introduction}

\defcitealias{2019ApJ...879....4W}{WT19}
\defcitealias{1980ApJ...236L.109L}{LE80}

Fast radio bursts (FRBs) are cosmological radio transients characterized by millisecond durations, large dispersion measures and implied brightness temperatures in excess of $10^{30}$ K. FRBs that repeat\footnote{also see \cite{1980ApJ...236L.109L}} are now well established \citep{2016Natur.531..202S,2019Natur.566..235C,2019ApJ...885L..24C,2020arXiv200103595F}, with the repeater FRB 121102 recently exhibiting tens of pulses within a span of hours \citep[e.g.,][]{2018ApJ...863....2G,2018ApJ...866..149Z,2019ApJ...877L..19G,2019arXiv191212217O}. It is an open question if the class of progenitors for FRBs which have been observed to repeat and those which have not (yet) repeated are identical. For recent reviews, see \citet{2018arXiv181005836P,2018PrPNP.103....1K, 2019A&ARv..27....4P}.
 
Given the pulsar-like (short duration, and high brightness temperature and linear polarization) properties of FRB pulses and large implied energies, flaring magnetars are a natural candidate for the progenitor \citep{2010vaoa.conf..129P,2013arXiv1307.4924P,2014MNRAS.442L...9L,2016ApJ...826..226K,2016MNRAS.461.1498M,2017ApJ...843L..26B,2017ApJ...838L..13L,2017MNRAS.468.2726K,2018ApJ...852..140W,2018ApJ...868...31Y,2019arXiv190103260L,2019MNRAS.485.4091M,2019ApJ...879....4W,2019arXiv190710394S,2020arXiv200102007L} with FRB emission transpiring within or external to the magnetosphere. {For instance, generic constraints on the emission mechanism in FRB 121102 by \cite{2017MNRAS.468.2726K} suggest magnetar-like fields may be involved in FRBs.} Magnetars in our galaxy, however, are known to undergo at least two distinct classes of X-ray and $\gamma$-ray flares: giant flares and short bursts. The former are more luminous with the initial spike much harder spectrally than short bursts, suggestive of initial deconfinement and outflows \citep[e.g.,][]{2016MNRAS.461..877V} with participation of large volumes of the magnetosphere and pair fireball creation \citep{1995MNRAS.275..255T,2001ApJ...561..980T}. Short bursts, in contrast, are far more numerous and two-blackbody fits are highly suggestive of trapping of plasma in the closed zones of magnetosphere involving a surface region a few percent of the NS surface area \citep[hot temperature component, e.g.,][]{2008ApJ...685.1114I,2010ApJ...716...97K,2012ApJ...756...54L,2012ApJ...749..122V,2014ApJ...785...52Y,2015ApJS..218...11C}. In epochs corresponding to these short burst episodes, quasi-periodic oscillations (QPOs) of frequencies $\nu \sim 20-300$ Hz have been reported -- these QPOs are interpreted as torsional eigenmodes of an oscillating crust \citep[e.g.,][]{2014ApJ...787..128H,2014ApJ...795..114H}. 

Recently, \cite{2019ApJ...879....4W} [\citetalias{2019ApJ...879....4W}] noted that the arrival time statistics of FRB 121102 are similar to magnetar short bursts {{\citep[see also][]{2019MNRAS.tmp.2746L}}}. \citetalias{2019ApJ...879....4W} proposed some FRBs are precipitated by field dislocations creating charge-starved regions in a low-twist (nearly curl-free) magnetosphere and lead to intense pair production if the preexisting charge density is sufficiently low. Pair cascades are a well-known required ingredient for radio emission in most pulsars {{\citep[e.g.,][]{1971ApJ...164..529S}}}. Logically, we assume FRBs can result if such avalanche pair cascades are produced in magnetars.  In this model, {\it{all such FRBs are associated with short bursts, but not all short bursts result in FRBs.}} Owing to the relatively low energy of short bursts, prompt time coincident high-energy counterparts to FRBs are not expected beyond a few Mpc \citep[e.g.,][]{2019ApJ...879...40C}. In \citetalias{2019ApJ...879....4W}, several conditions must also be satisfied for the operation of a putative pulsar-like emission mechanism. These requirements are readily satisfied in low-twist magnetars undergoing torsional crustal oscillations dislocating magnetic foot points. Yet, these conditions may not be satisfied for objects with lower fields or shorter periods, such as ``dead neutron stars" thought to significantly outnumber the observed pulsar/magnetar population, or much younger and short-lived millisecond magnetars. 

{The specifics of pulsar-like polar cap radio emission mechanism(s) invoked in this work for recurrent FRBs is beyond the scope of this paper. The problem of pulsar radio emission mechanism has proven to be notoriously complicated -- there is still no satisfactory model for it \citep[see][]{2017RvMPP...1....5M}. The most popular models invoke radiation of particle bunches \citep[e.g.,][]{1975ApJ...196...51R} or two-stream instability in relativistic pair plasmas \citep[e.g.,][]{1987ApJ...320..333U}. In either of these approaches a stationary plasma flow is assumed, with relativistic plasma streaming orderly in a single direction, generally away from the neutron star (NS). However, recent self-consistent local plasma simulations of magnetic pair cascades for pulsar polar caps \citep[][]{2010MNRAS.408.2092T,2013MNRAS.429...20T} have demonstrated that plasma flow is highly non-stationary and non-unidirectional with no evidence of charge bunches. Therefore, it highly unlikely that any of the several variants of existing radio emission models are realistic. On the other hand, electron-positron plasma formation fronts in polar caps is produced in ``discharges", when freshly-created pair plasma screens {\textit{in situ}} the particle-accelerating electric fields. These discharges are accompanied by large amplitude fluctuations of electric field and collective plasma motion. It would be reasonable to assume that such discharges can generate coherent electromagnetic radiation. Recently \cite{2020arXiv200102236P} demonstrated that such discharges can indeed excite superluminal electromagnetic plasma waves that could eventually escape the magnetosphere. The details of this novel coherent radio emission mechanism have yet to be worked out, so it would be premature to make any prediction about the efficiency of this mechanism for our FRB context. Therefore, here we simply postulate that magnetic pair production in NS magnetospheres is a necessary ingredient for coherent polar-cap radio emission.}

In this paper, we explore the viable region of the NS $P-\dot{P}$ parameter space for {{recurrent}} FRBs in the low-twist magnetar model. We assume field dislocations (or crust oscillations) occur with a single characteristic frequency $\nu$ and wavelength $\lambda$, with varying spatial displacement amplitude $\xi$, i.e. we only consider a single dominant eigenmode. The details and rich complexity of NS crust physics and how the internal/external fields couple to the crust, including anisotropy of the strain tensor and the power spectrum of excited higher harmonics, is not well understood \citep[for a review, cf.][]{2008LRR....11...10C}. Moreover, it may not alter our principal conclusions if a preferential narrow regime of eigenmodes is demanded for FRB production, or if one fundamental eigenmode dominates excitations. For this first exploration, we adopt a minimalist point of view and defer such complexity to the future.

{{For FRB repeaters,}} we formulate an FRB death line in the traditional $P-\dot{P}$ diagram, adopting simple necessary criteria for charge starvation and magnetic dominance in \S\ref{constraints}. The FRB NS hosts may not lie past the death line in a region of low magnetic field $B$ and short spin period $P$. Motivated by the power law fluence distribution of FRB 121102, we propose an energy/luminosity function of FRBs for individual repeaters as a function of the  oscillation amplitude in \S\ref{reverseengineering} via an ``inversion protocol". We apply this elementary model to an evolved population of magnetars, proposing observationally-constrained models of magnetic field decay for magnetars in our galaxy, to obtain population luminosity functions of FRBs for different potential FRB host populations in \S\ref{populationE}. Assuming all FRBs arise from low-twist magnetars, we also compute the local fluence distribution in standard cosmology for different potential subpopulations of magnetars. An extensive discussion with an observational focus follows in \S\ref{summarydiscussion}.

\newpage
\section{The Model: Constraints on Neutron Star Period and Surface Magnetic Field for FRB Viability}
\label{constraints}

In the model proposed by \citetalias{2019ApJ...879....4W}, field dislocations and oscillations at the NS surface induce strong electric fields. There are three relevant magnetospheric charge densities, which characterize the values that are capable to screen accelerating electric fields induced by (i) corotation \citep[the Goldreich-Julian charge density,][]{1969ApJ...157..869G} $\rho_{\rm corot}$, (ii) field twist $\rho_{\rm twist}$, and (iii) field dislocations $\rho_{\rm burst}$. If the corotation or field twist charge density present near the surface prior to the field dislocation is insufficient to screen the burst induced electric field, then intense electron/positron pair cascades can result on the timescale of microseconds associated with acceleration gaps. The burst electric field timescale may set the characteristic duration of FRB pulses; the particle acceleration may persist until the flux tube is crowded by plasma and the global electric field screened, or for the field dislocation timescale $\Delta t$. Moreover, since charge densities and associated plasma frequencies are low throughout the magnetosphere, the low-twist state facilitates escape of radio waves. The model is very similar to that of polar cap cascades thought to underlie coherent radio emission in most radio-loud pulsars and magnetars, but differentiated by the origin and transient nature of the particle acceleration. Frequency drifts observed in many recurrent FRBs \citep[e.g.,][]{2019ApJ...876L..23H,2019ApJ...885L..24C}, hitherto universally nonpositive, could result from pulsar-like radius-to-frequency mapping, with drifts owing to the decline of plasma density (and plasma frequency) with altitude in a flux tube as noted in \citetalias{2019ApJ...879....4W} and also \citep{2019arXiv190103260L,2019arXiv190807313L}.

Note that in this model we are agnostic as to whether the trigger is internal \citep[e.g.,][]{2011ApJ...727L..51P,2013MNRAS.434..123V,2019arXiv190710394S} or external \citep[e.g.,][]{2015MNRAS.447.1407L}; provided that field dislocations are coupled to the crust,  such dislocations are charge-starved and $B$ is locally large enough for magnetic pair cascades. Even in magnetospheric reconnection models where the energy release is external, the field ought to be coupled to the electrons in the crust, to account for the recently inferred QPOs during short bursts \citep[e.g,][]{2014ApJ...787..128H,2014ApJ...795..114H}. In either internally or externally-triggered scenarios, a scaling of the square amplitude of the crust/field oscillations (in the elastic regime) may be regarded as a {{\it proxy}} for the event's energy release (see below and \S\ref{confinement}). Additionally, since the core-crust coupling is suggested by the damping of QPOs within $\lesssim 0.2-2$ s in short bursts \citep{2014ApJ...793..129H,2019ApJ...871...95M}, such damping may limit the duration of FRB recurrence clusters (that is, the span of short-waiting-time {trains}, as exhibited in FRB 121102) for individual triggering events. Multiple field dislocation episodes then correspond to similar FRB repetition statistics as magnetar short bursts. 

Summarizing, in the context of \citetalias{2019ApJ...879....4W}, there are three necessary (but not sufficient) requirements for the viability of FRBs from NSs undergoing field dislocations. 

\begin{itemize}
\item First, prior to the burst, the local charge density in the magnetosphere linked to the burst active region must be sufficiently low so that the burst Goldreich-Julian density $\rho_{\rm burst}$ is larger, i.e. field oscillations should cause charge starvation. This condition is necessary for particle acceleration and avalanche pair production precipitated by dislocations of magnetic foot points, and sets a maximum corotation Goldreich-Julian charge density $\rho_{\rm corot}$ that may exist prior to a burst of specified amplitude, as extensively discussed in \citetalias{2019ApJ...879....4W}.

\item Second, we propose that the plasma remain magnetically dominated during bursts. Since the open zone is small in slow rotators, such bursts would predominantly involve magnetic foot points which tap closed zones of the magnetosphere. Confinement of plasma in magnetic flux tubes is suggested by two-blackbody phenomenology in magnetar short bursts \citep[e.g.,][]{2014ApJ...785...52Y}. Strong and ordered magnetic fields are also indicated in FRB 121102 by its high linear polarization and fixed polarization angle (PA) during bursts \citep{2018Natur.553..182M}. Moreover, the pulsar-like emission mechanism itself may require a magnetically-dominated plasma for its operation, since the generation of coherent plasma waves thought to be a necessary ingredient in its operation. Such magnetic dominance is certainly realized for polar cap zones where radio emission is thought to originate in most pulsars, but is much more constraining for the proposed FRB model.

\item Third, the magnetic field must be curved\footnote{Curvature is a necessary condition for magnetic pair production which requires nonzero photon angles to the field. This is automatically guaranteed in dipolar morphologies.} and high enough for prolific magnetic pair production to arise during a burst of specified amplitude $\xi$, a situation readily true for magnetars and high-B pulsars but not objects with surface fields below, as we show, about $10^{11}$ G, for amplitudes $\xi$ compatible with charge starvation. 
\end{itemize}

Above, it is understood that the NS is susceptible to magnetar-like short bursts via magnetoelastic crust deformations over a large range of amplitudes/energies. This prerequisite depends on the formation of a magnetized crust in conditions of high pressures and $B$ fields \citep[e.g.,][]{2001RvMP...73..629L,2006PhRvA..74f2508M,2006RPPh...69.2631H}, a physical regime that is generally poorly constrained. In either hydrogen \citep{1997ApJ...491..270L} or higher $Z$ element \citep{2007MNRAS.382.1833M} NS envelops, it has been suggested that there exists a critical temperature $T_{\rm crit} \lesssim 10^7$ K below which a condensed matter phase may form in high $B \sim 10^{15}$ G fields. This condensed phase may also facilitate the formation of accelerating gaps during field dislocations. The critical temperature is generally too low for condensed crusts to be plausible for newborn millisecond magnetars, but is above the typical surface temperatures inferred from thermal persistent emission in known aged magnetars and high-B pulsars. In \S\ref{populationE} we circumvent this uncertainty of crust physics by adopting $\{P,\dot{P}\}$ distributions corresponding to the known galactic population of magnetars assuming dipole spin-down\footnote{This is a lower limit on surface $B$, and therefore a conservative estimate \citep[multipolar components may exist, e.g.,][]{2013Natur.500..312T}.} $B \sim 6.4 \times 10^{19} \sqrt{P \dot{P}}$ G.

In this study, we assume the NS field twist is sufficiently low (globally or in localized regions of the magnetosphere) so that the {\it{corotational Goldreich-Julian charge density}} is the limiting factor, i.e. $\rho_{\rm twist} \sim [c B/(4 \pi R_*)] \sin^2 \theta_0 \Delta \phi  \lesssim \rho_{\rm corot} $ or $\Delta \phi \lesssim 4 \pi R_*/(c\, P\, \sin^2 \theta_0) \sim 4 \times 10^{-5} (P/10 \rm \,\, s)^{-1} \sin^{-2} \theta_0$ where $\Delta \phi$ is the dimensionless twist, $R_* \sim 10^6$ cm is the radius of the NS, and $\theta_0$ is a magnetic foot point colatitude (see \citetalias{2019ApJ...879....4W} for details). This twist angle is much smaller than that assumed in standard magnetar models for nonthermal persistent emission \citep[e.g.,][]{2007ApJ...657..967B}. That is, in this paper we consider the space of all possible magnetars which could produce FRBs if a state of low twist concomitant with field dislocations conditions could occur. This larger set may be sieved further by other physical factors that select which magnetars are in a state of low twist or can undergo strong field dislocations that yield FRBs. 

\paragraph{Characteristic Scales} During crustal oscillations, the value of maximum amplitude $\xi_{\rm max}$ is set by the characteristic yielding strain of the crust $\sigma_{\rm max}$ where $\sigma \equiv \Delta x/x \sim \xi/\lambda$, $\lambda \lesssim R_*$ and $\xi_{\rm max} = \lambda \sigma_{\rm max}$. Below this maximum strain, we assume torsional magnetoelastic oscillations are viable, and under our assumptions, FRBs production is tenable if other conditions, detailed below, are met. Since the characteristic shear wave speed in the crust is estimated to be of order $v_s \sim {\rm few} \times 10^7$ cm s$^{-1}$ \citep[e.g.,][]{2009PhRvL.103r1101S}, for observed QPO frequencies of $\nu\sim100$ Hz and since the wavelength is $\lambda \sim 10^{5.5}$ cm, this implies $\xi_{\rm max} \lesssim 10^{4.5}$ cm for yielding $\sigma_{\rm max} \sim 0.1$ strain. Recent studies support a high value of $\sigma_{\rm max} \sim 0.1$ \citep{2009PhRvL.102s1102H,2012MNRAS.426.2404H}, and as will be apparent, this is also what we find more favorable for FRB production. Events corresponding to single dislocations events of duration $\Delta t$ may be regarded as $\nu \leftrightarrow 1/\Delta t$; if such dislocations excite higher harmonics or overtones, $\Delta t$ may be much shorter than the typical oscillation period inferred in known magnetars.

\subsection{Local Charge Starvation}
\label{chargestarvation}

Given sufficiently low twist in the active region, the charge starvation condition (Eq.~(5)) from \citetalias{2019ApJ...879....4W} takes the form,
\begin{equation}
\rho_{\rm burst} > \rho_{\rm corot}.
\end{equation}
Here, $\rho_{\rm burst}$ is the charge density required to screen the electric field induced by field dislocations from Eq.~(3)--(4) of \citetalias{2019ApJ...879....4W}. Then, 
\begin{equation}
\rho_{\rm burst} \sim \frac{1}{2} \frac{\xi}{\lambda} \frac{\nu}{c} B \quad , \quad \rho_{\rm corot} \sim \frac{B}{c \, P}
\label{rhoburst}
\end{equation}
implies a minimum period independent of surface $B$,
\begin{equation}
P \gtrsim   P_S \equiv \frac{2 \lambda}{\nu \xi} > P_{S,\rm min} = \frac{2}{\nu \sigma_{\rm max}} \approx  0.2 \, \left(\frac{\nu}{100 \, \rm Hz} \right)^{-1}  \left( \frac{\sigma_{\rm max}}{0.1} \right)^{-1} \quad \rm sec.
\label{Pstarve}
\end{equation}
Equivalently for a given period, the charge starvation condition requires,
\begin{equation}
\nu > \frac{2}{P \sigma_{\rm max} }
\label{minnu}
\end{equation}
which manifestly couples the FRB-viable frequency of oscillations to the crust breaking strain. Clearly, the existence of a minimum period implies a minimum age for young FRB progenitors in this model. The characteristic scale of $P_{S,\rm min}$ above also largely rules out recycled millisecond NSs. We note that too small a value of $\xi$ would lead to implausibly long periods, which are not realized in most observed magnetars, so this effectively sets a threshold on $\xi$ when $\lambda,\nu$ are fixed, i.e., 
\begin{equation}
\xi \gtrsim \xi_{\rm min} = \frac{2 \lambda}{\nu P} \approx 6 \times 10^2 \frac{\lambda_{5.5}}{\nu_{2} P_{1}}\quad \rm cm
\label{ximinP}
\end{equation}
where hereafter we adopt the notation ${\cal X}_y   \equiv {\cal X} \, 10^{-y}$ in cgs units.

\subsection{Constraints by Magnetic Pair Cascade Viability}

In Appendix \ref{pairviability} we demonstrate that relatively low local fields of $B \gtrsim 10^{11}$ G are required to initiate single photon magnetic pair cascades during the smallest viable field dislocation amplitudes Eq.~(\ref{ximinP}). These fields, however, limit the altitude of where pulsar-like pair cascades may occur to about a few stellar radii for typical magnetar dipole field components. For magnetar fields in the absence of photon splitting, gap heights are of order $h_{\rm gap} \sim 10^{3}-10^4$ cm (Eq.~(\ref{hgapthres})) and regulated by curvature cascades near pair threshold. As detailed in Appendix \ref{splitting}, photon splitting is generally not expected to quench pair cascades (and therefore radio emission in the proposed model) even if all photon polarization state channels allowed by CP symmetry operate in most of the magnetopshere.

\subsection{Magnetic Confinement}
\label{confinement}

In known magnetars, since high-energy short bursts are quasi-thermal, they are calorimetric for the event energy release. The short burst energies span a large range ${\cal E}_{\rm sb} \sim 10^{36}-10^{42}$ erg. As in earthquakes or solar flares, the number energy/fluence distribution of magnetar short bursts are known to follow a power law distribution, $dN/d{\cal E}_{\rm sb} \propto {\cal E}_{\rm sb}^{-s}$, typically with $s \approx 1.6-1.7$ across many magnetars and episodes of individual magnetars \citep{1996Natur.382..518C,1999ApJ...526L..93G,1999ApJ...519L.139W,2000ApJ...532L.121G,2004ApJ...607..959G,2010A&A...510A..77S,2011ApJ...739...94S,2012ApJ...755....1P,2012ApJ...749..122V,2015ApJS..218...11C}. Although deviations from $s \sim 1.6-1.7$ have been reported \citep[e.g.,][]{2001ApJS..137..227A,2011ApJ...739...87L}, $s < 2$ appears universal, implying the largest events dominate the energetics.  

As noted earlier, we are agnostic to whether the short burst trigger is internal or external provided that the field couples to the crust. The energy of ${\cal E}_{\rm sb, max}$ is related to the mass participating in the crustal oscillations via field/crust dislocation $\xi$. In the elastic regime, the energy scales as the square of the amplitude: $E_{\rm elas} \sim (1/2) \Delta M \nu^2 \xi^2$ where $\Delta M \sim (\lambda/R_*)^2 M_{\rm C}$ with $M_{\rm C}$ the mass of the Coulomb lattice in the crust \citep[e.g.][]{2001ApJ...561..980T}. The factor $ (\lambda/R_*)^2$ represents the fraction area of the active region and accounts for the detail that the hotter area in short burst two-blackbody fits indicates a surface active region of a few percent (rather than the whole) of the NS surface area.  If the largest ${\cal E}_{\rm sb, max}  \sim 10^{42}$ erg \citep[e.g.,][]{1996Natur.382..518C,2012ApJ...749..122V} events occur at maximum amplitude $\xi_{\rm max}$, this implies $\xi_{\rm  min} \lesssim 10^{-3} \xi_{\rm max}$ for the lower energy events of ${\cal E}_{\rm sb} \sim 10^{36}$  erg. 

Since $\xi_{\rm max} = \lambda \sigma_{\rm max}$ and $\nu \lambda \sim v_s$ we obtain the parameterization, 
\begin{equation}
{\cal E}_{\rm sb} \sim \eta E_{\rm elas} \sim \frac{1}{2} \eta \sigma_{\rm max}^2 M_{\rm C} v_s^2 \left(\frac{\lambda}{R_*}\right)^2 \left(\frac{\xi}{\xi_{\rm max}}\right)^2  = \eta E_{\rm max} \left(\frac{\xi}{\xi_{\rm max}}\right)^2   \approx 3 \times 10^{43} \, \eta \left(\frac{\xi}{\xi_{\rm max}}\right)^2   \quad \rm erg
\label{Eelas}
\end{equation}
where we have assumed $M_{\rm C} = 0.03 M_\odot$, estimated in \cite{2001ApJ...561..980T} \citep[also see \S6 of][]{2008LRR....11...10C} and $E_{\rm max} \sim \sigma_{\rm max}^2 M_{\rm C} v_s^2 (\lambda/R_*)^2/2$. We list adopted values for other parameters in Table~\ref{tab:paras}. Here $\eta \sim \mathcal{O}(0.1)$ is the conversion efficiency into observed X-rays as inferred from the largest observed short bursts of ${\cal E}_{\rm sb, max} \sim 10^{42}$ erg. In either internal \citep[e.g.,][]{2015MNRAS.449.2047L} or external energy release scenarios, $E_{\rm max} $ could be considerably larger than the adopted conservative value of $10^{43.5}$ erg. To remain agnostic, we regard $E_{\rm max} $ as a parameter in our model.

For FRBs resulting from such field dislocations, we propose the plasma at the base of the flux tube is similarly magnetically dominated, in order to provide magnetic confinement at higher altitudes, i.e. $B^2/(8 \pi) \gtrsim  E_{\rm elas}/(\lambda^2 \Delta R_*)$ at the surface\footnote{If nonideal plasma processes and reconnection are unimportant, then, owing to flux freezing the plasma density drops proportional to B at higher altitudes.} where $\Delta R_* \sim 0.1 R_* \approx 10^5$ cm is the crust thickness, a quantity which captures the characteristic depth of the active region which the internal magnetic field threads. From the expression for $ E_{\rm elas}$ in Eq.~(\ref{Eelas}), we arrive at

\begin{equation}
B \gtrsim  B_{\rm mag} (\xi) = \frac{\xi}{\xi_{\rm max} } \left( \frac{8 \pi {E}_{\rm  max} }{\lambda^2 \Delta R_* } \right)^{1/2}  \approx   3 \times 10^{14}  \,  \, \frac{\xi}{\xi_{\rm max}} \left( \frac{ {E}_{\rm max, 43.5}}{ \lambda_{5.5}^2 \Delta R_{*,5}} \right)^{1/2}  \quad \rm  G.
\label{Bmag}
\end{equation}
If $B < B_{\rm mag} (\xi_{\rm max})$, then the maximum amplitude which satisfies magnetic dominance is
\begin{equation}
\xi \lesssim \xi_{\rm conf} = B \lambda^2 \sigma_{\rm max} \sqrt{\frac{\Delta R_*}{8 \pi {E}_{\rm  max}}} \sim 10^{5} B_{15} \quad \rm cm
\end{equation}
for adopted parameters in Table~\ref{tab:paras}.

\begin{deluxetable*}{l|c|c}
\tablenum{1}
\tablecaption{Adopted Values for Parameters \label{tab:paras}}
\setlength\tabcolsep{5mm}
\tablehead{
\colhead{Quantity} & \colhead{Variable} & \colhead{Adopted Values} }
\startdata
Index of short burst energies/fluences $dN/d{\cal E}_{\rm sb} \propto {\cal E}_{\rm sb}^{-s}$ & $s$ & 1.7 \\
Characteristic crust/field oscillation wavelength & $\lambda$ & $10^{5.5}$ cm \\
Characteristic crust/field oscillation frequency & $\nu$ & $100$ Hz \\ 
Crust breaking strain & $\sigma_{\rm max}$ & $0.1$ \\
Characteristic crust thickness & $\Delta R_*$ & $10^5$ cm \\
Maximum crust/field dislocation amplitude $\lambda \sigma_{\rm max}$ & $\xi_{\rm max}$ & $10^{4.5}$ cm \\
Event energy release at $\xi_{\rm max}$ & $E_{\rm max}$ & $10^{43.5}$ erg \\
\hline
FRB isotropic-equivalent energy and field scale factors [erg, G]  & $[{\cal E}_0, B_r ] $ & \parbox[t]{1.5cm}{$[10^{40}  , 10^{15} ]$ \\ $[10^{41}  , 10^{15} ]$ \\ $[10^{42} , 10^{15}]$ } \\
FRB individual repeater energy function $\xi B$ exponent & $a$ & 1 \\
\hline
Birth magnetic field range & $B_0$ & \parbox[t]{3.5cm}{I: $3\times 10^{14}- 10^{15}$   G \\ II: $3\times 10^{14}- 3\times 10^{15}$  G} \\
Index of magnetic field evolution & $\alpha$ & $\{ -1,0,1\}$ \\ 
Timescale of B decay & $\tau_B$ & $10^4$ yr
\enddata
\tablecomments{Multiple values for adopted parameters ($\alpha$, $B_0$ and $[{\cal E}_0, B_r ] $) are subsequently employed in Figures~\ref{ppdotdiagram}--\ref{localfluencedist}.   }
\end{deluxetable*}

\subsection{The FRB Death Line}

\begin{figure}[t]
\centering
\includegraphics[width=0.999\textwidth]{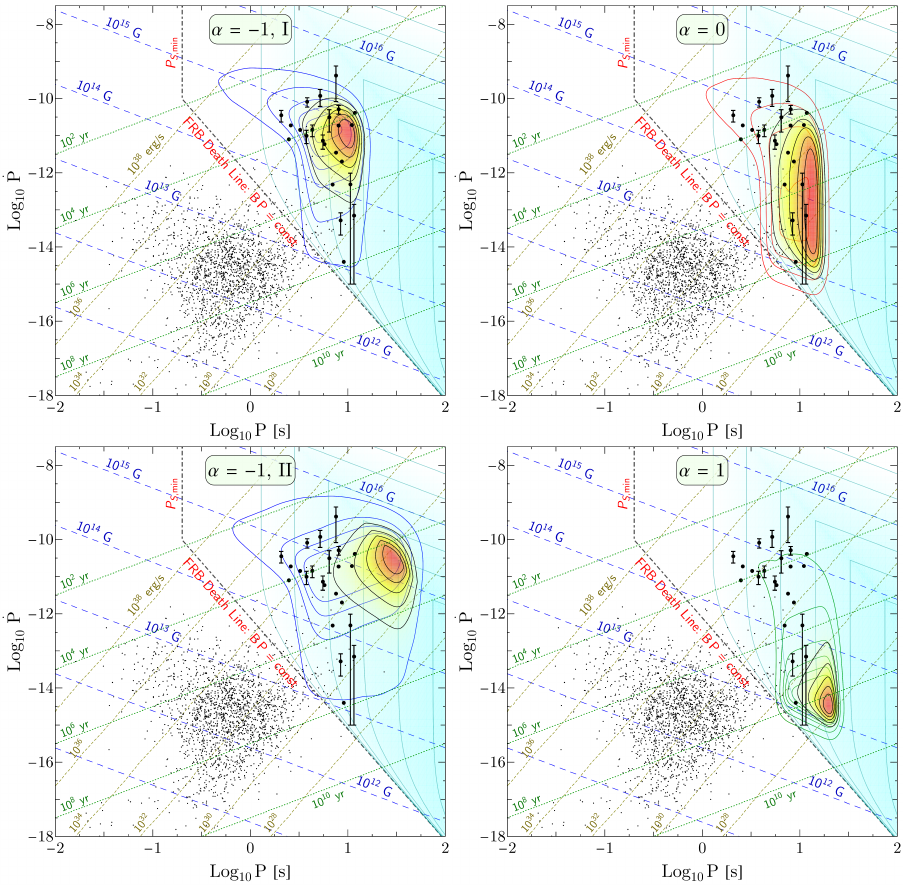}
\caption{$P-\dot{P}$ diagrams illustrating the viable space of NSs which may produce FRBs, with panels of different magnetic evolution parameter $\alpha$. Lines for constant characteristic age $P/\dot{P}$, dipole-component magnetic field $B=6.4 \times 10^{19} \sqrt{P \dot{P}}$~G and spin-down power are illustrated along with the pulsar population from the ATNF catalog \citep{2005AJ....129.1993M}.  In addition, the known galactic magnetar population is depicted with heavy weighted points along with uncertainties on $\dot{P}$. Contours reflecting the probability density of the magnetar population from \cite{2019MNRAS.487.1426B} are illustrated in $\rm \{blue, red, green \}$ corresponding to $\alpha = \{-1,0,1\}$ respectively. For $\alpha = \pm 1$, successive colored contours depict the normalized density of evolved magnetars at $[0.01, 0.1,0.3,0.5,0.7,0.9]$ level of the maximum; for $\alpha =0$, the red contour levels are set at $[0.01, 0.1,0.3,0.5]$. Contours depicting the acceptance fraction $A_f$ in $P-\dot{P}$ space are in dark cyan for $[{\cal E}_0, B_r ] = [10^{41} {\rm \, erg} , 10^{15} {\rm  \, G} ]$, the adopted burst amplitude distribution and other parameters listed in Table~\ref{tab:paras} and subject to all constraints mentioned in the text. Low values of $A_f$ asymptote to the death lines depicted with the black dot-dashed lines. The upper left death line is set by the adopted maximum amplitude in the model corresponding to a minimum period in Eq.~(\ref{Pstarve}) while the the other line follows from magnetic confinement with charge starvation, Eq.~(\ref{deathlinesimp}). The constraints associated with feasibility of pair cascades (\S\ref{pairviability}) is off the scale. The intersection distribution convolving the acceptance fraction and magnetic population is depicted with yellow-red colors and five black contours spanning factor of five in density. For $\alpha = -1$, ``normal" (I) and ``extended" (II) birth $B_0$ models for populations are depicted.}
\label{ppdotdiagram}
\end{figure}

At equality, the conditions $B \geq B_{\rm mag}$ and $P \geq P_S$ constitute a death line in the $P-\dot{P}$ diagram due to the combination of magnetic confinement and charge starvation. The product of $B$ and $P$ is of the independent the oscillation amplitude $\xi$ (obtained by eliminating $\xi$ in Eqs.~(\ref{Pstarve}) and (\ref{Bmag})):
\begin{equation}
B \, P \gtrsim  \frac{\lambda}{R_*} \sqrt{\frac{16 \pi M_{\rm C}}{\Delta R_*} } =  \sqrt{ \frac{32 \pi {E}_{\rm max}}{\Delta R_* \xi_{\rm max}^2 \nu^2} }\sim 6 \times 10^{13} \, \, {E}_{\rm max,43.5}^{1/2} (\Delta R_{*,5})^{-1/2} \xi_{\rm max,4.5}^{-1} \nu_{2}^{-1} \quad \rm G \, \, s
\label{deathlinesimp}
\end{equation} 
which manifestly segregates the magnetars and high-B pulsars from the bulk of the NS population -- see Figure~\ref{ppdotdiagram}. $B$ and $P$ in Eq.~(\ref{deathlinesimp}) span from the minimum value $P = P_{S,\rm min}$ to a minimum value of surface $B = B_{\rm gap} (\xi_{\rm max})$ which supports pair cascades or one that is consistent with the existence of a magnetized crust.

\section{The Radio Burst Energy Distribution for Individual Magnetars via an Inversion Protocol}
\label{reverseengineering}

For FRB 121102's bursts reported by \cite{2018ApJ...866..149Z},  the burst fluence distribution of FRB 121102 exhibits a power law where \citetalias{2019ApJ...879....4W} find $dN/d{\cal F} \propto {\cal F}^{-\Gamma}$ with $\Gamma \sim 2.3 \pm 0.2$ using a maximum likelihood technique for ${\cal F} \sim 30 - 600$ Jy $\mu$s. For a lower energy sample of Arecibo bursts at $\nu_r \sim 1.4$ GHz, \cite{2019ApJ...877L..19G} report $\Gamma \sim 2.8 \pm 0.3$. {Likewise, \cite{2019arXiv191212217O} independently report $\Gamma \sim 2.3-2.7 $ for FRB 121102 bursts observed by WSRT/Apertif.} Yet, although $\Gamma$ is rather uncertain, the power law or heavy-tailed nature rather than exponential character seems robust as well as its greater steepness over that of short bursts, $\Gamma > 2 > s$.
 
Although the radio emission mechanism(s) in NS magnetospheres is poorly understood, in the context of our model where the same trigger underpins short bursts and FRBs, the power-law distribution of FRB fluences encourages coupling the FRB energy directly with the amplitude $\xi$. That is, the fluence distribution of both FRBs and short bursts is assumed to arise from the same amplitude distribution that emerges from the dynamics and condensed matter physics of the crust or reconnection within the magnetosphere which couples to the crust \citep[e.g.,][]{2015MNRAS.447.1407L}. 

Yet, unlike the high-energy short bursts that scale as the square of the amplitude, the FRB dependence on amplitude ought to contain a $B$ dependence since the screening condition is $\rho_{\rm burst} \propto \xi B$ in our model. Consequently, we propose a simple heuristic model where the FRB energy scales as a power of the duo $B$ and $\xi$:
 \begin{equation}
{\cal E}_r  \equiv {\cal E}_{0} \left( \frac{\xi}{\xi_{\rm max}} \frac{B}{B_r}  \right)^a  \,  \Theta(\nu, \lambda, \xi, P, \dot{P}) \qquad , \qquad a < 2
\label{Efunc}
\end{equation}
where $\Theta$ imposes cutoffs in the luminosity function when the conditions for charge starvation, magnetic dominance, particle acceleration and pair production are not met. The factor $ {{\cal E}_{0} \sim 10^{40}-10^{42}}$ erg is the empirically determined isotropic-equivalent cutoff energy of the FRB luminosity function \citep[e.g.,][]{2018MNRAS.481.2320L}. For $\Theta$, we adopt the form
 \begin{equation}
 \Theta(\nu, \lambda, \xi, P, \dot{P}) \; =\;  
 \begin{cases}
       1 & \mbox{if } \quad  B > \max\{ B_{\rm gap}, B_{\rm mag}\} \, \,  \textrm{and} \, \,  P > P_S \, \,  \textrm{and}\, \,   {\cal E}_r < {\cal E}_{\rm sb}  \\
       0 & \mbox{otherwise}.
  \end{cases}
  \label{Thetafunc}
 \end{equation}
The inequalities effectively set lower and higher cutoffs on the viable amplitudes. From the charge starvation in WT19, we have $\xi > 2 \lambda/(\nu P)$ (see \S\ref{chargestarvation}). We see that, in our model, there is intrinsically a lower energy scale, ${\cal E}_{r, \rm min}$, in the energy distribution for a single magnetar.

With the assumption ${\cal E}_{\rm sb} \propto \xi^2$ (Eq.~(\ref{Eelas})), the distribution of $\xi$ also follows a power law:  $dN/d\xi \propto \xi^{-(2 s-1)}$ over $\xi \in \{\xi_{\rm min}, \xi_{\rm max} \} \sim \lambda \sigma_{\rm max} \{x, 1 \}$ with $x \ll 1$ which captures the operating of amplitude fluctuations. We will see below that the precise value $x$ is irrelevant when it is sufficiently small. The normalized PDF of amplitude magnitudes is,
\begin{equation}
\frac{dN}{d \xi} = \left(\frac{2s - 2 }{\left[\lambda \sigma_{\rm max}\right]^{2-2s} \left(x^{2-2s}-1 \right) } \right) \xi^{-(2 s -1)} \qquad  x < \xi < \lambda \sigma_{\rm max}.
\label{xipdf}
\end{equation}
We can obtain the energy distribution $dN/d{\cal E}_r$ by drawing $\xi$ from Eq.~(\ref{xipdf}) assuming the relation Eq.~(\ref{Efunc}) between ${\cal E}_r$ and $\xi$. When B is sufficiently large and magnetic dominance is maintained $(\Theta\rightarrow1$), the distribution of burst energies is a power law (with boundaries given by Eq.~(\ref{Ecuts}) below),
\begin{equation}
\frac{dN}{d{\cal E}_r}  = \frac{dN}{d\xi} \left(\frac{d{\cal E}_r}{d\xi} \right)^{-1} \propto {\cal E}_r^{-(2s +a-2)/a}.
\label{dndEr}
\end{equation}
Then, $\Gamma = (2s +a-2)/a$ which implies 
\begin{equation}
a = \frac{2 (s-1)}{\Gamma - 1} \approx  1.08  \qquad   {\rm when} \quad s \approx 1.7 {\quad \rm and \quad} \Gamma \approx 2.3 
\label{avalue}
\end{equation}
This intriguing value of $a \sim 1$ suggests ${\cal E}_r \propto \xi B$, i.e. the FRB energy directly scales with the voltage drop or primary particle energy in gaps, i.e. Eq.~(\ref{Egap})--(\ref{gammaacc}). This is a result reminiscent to that obtained empirically for radio pulsars by \cite{2002ApJ...568..289A}, and suggests that beaming or propagation influences may be minor for the inferred energies. This curious correspondence also supports the assumption in this work that a magnetospheric pulsar-like emission mechanism underpins recurrent FRBs. 

The radio energy cannot exceed the event energy budget, ${f_b} {\cal E}_r < E_{\rm elas}$ where ${f_b} < 1$ is a beaming correction factor accounting for anisotropy of radio emission for the isotropic-equivalent ${\cal E}_r$. This demand also sets a lower cutoff $\xi_{\min}$ such that  ${f_b} {\cal E}_r (\xi)/ E_{\rm elas}(\xi) < 1$, yielding,
\begin{equation}
\xi > \xi_{\rm min} \equiv \xi_{\rm max} \left( \frac{{f_b} {\cal E}_0}{ E_{\rm max}} \right) ^{\frac{1}{2-a}} \left( \frac{B}{B_r} \right) ^{\frac{a}{2-a}}.
\label{efficiencyconstraint}
\end{equation}
Since $\xi/\xi_{\rm max} \leq 1$, there is an upper bound on $B$, which is largely immaterial for chosen parameters since is it is beyond the range plausible for stability in NSs,
\begin{equation}
B \ll B_r \left( \frac{E_{\rm max}}{{f_b} {\cal E}_0} \right)^{1/a} \sim 10^{17} - 10^{18} \qquad \rm G.
\end{equation}
Let us now consider the maximum burst energy bounds in the adopted model, realized for $B$ sufficiently large so that magnetic dominance is maintained. Explicitly, the minimum and maximum radio burst energies are 
\begin{equation}
\left\{ {\cal E}_{r, \rm min}, {\cal E}_{r, \rm max} \right\} = {\cal E}_{0}\left( \frac{B}{B_r}\right)^a \left\{ \max \left[ \epsilon_1 , \,  \epsilon_2 \right], \, 1 \right\}.
\label{Ecuts}
\end{equation}
where 
\begin{eqnarray}
\epsilon_1 &=& \left( \frac{2}{\nu P \sigma_{\rm max}} \right)^a  \nonumber \\
 \epsilon_2 &=& \left( \frac{B}{B_r} \right)^{a^2/(2-a)} \left( \frac{{f_b} {\cal E}_0}{{E}_{\rm max}} \right)^{\frac{a}{2-a}} .
 \label{epsilonEcuts}
\end{eqnarray}
These $\epsilon_1$ and $\epsilon_2$ arise from charge starvation and the adopted energy condition  Eq.~(\ref{efficiencyconstraint}), respectively. Hereafter, we adopt the beaming factor ${f_b} =1$ and regard all quantities as ``isotropic equivalent" over a large statistical ensemble of bursts, omitting frequency-dependent selection effects. In our model, such a cutoff energy scale arises naturally if magnetic dominance is required for the FRB process in short bursts of energy ${E}_{\rm  max} \sim 10^{43} - 10^{44}$ erg in magnetar-like fields of $B \sim 10^{15}$ G (see Eq.~(\ref{Bmag})), and if the conversion efficiency of $E_{\rm max}$ into an FRB is appreciably less than unity.  For typical parameter values, $\epsilon_1 \gg \epsilon_2$, so that the charge starvation condition is the limiting factor for the smallest amplitudes corresponding to the lowest energy bursts.

When $B$ is not large enough in Eq.~(\ref{Bmag}) to confine a burst with amplitude $\xi < \xi_{\rm max}$, the $\Theta$ function imposes a high energy cutoff, further narrowing the range of realizable fluences. Such a dynamic range is also further restricted when there is a preexisting low field twist where $\rho_{\rm burst} > \rho_{\rm twist} > \rho_{\rm corot}$ so that the minimum burst amplitude to overcome this density is larger than simply $\rho_{\rm corot}$ (see \S\ref{finitetwist}). Hence, the span of energies (or fluences) of individual repeaters is inherently quite narrow in the adopted model. Then, assuming $B$ is sufficiently large to satisfy $B > \max\{ B_{\rm gap}, B_{\rm mag} \}$ the range Eq.~(\ref{Ecuts}) constitutes the maximum possible range of burst energies. Here, different values of $2/(\nu P \sigma_{\rm max})$ alter the span of the burst energy distribution, while $B$ translates the entire distribution to lower or higher energies. We see that this dynamic range is independent of $x$. For FRB 121102, the span of burst fluences is at least $1.5$ orders of magnitude, so $2/(\nu P \sigma_{\rm max}) \lesssim 10^{-1.5}$. For $P\sim 10$ s and $\nu \sim 100$ Hz, this implies $\sigma_{\rm max} \gtrsim 0.06$. Alternatively, for a breaking strain of $\sigma_{\rm max} \sim 0.1$, we see the fluence distribution likely does not exceed a span of $10^3$ for oscillation frequencies $\nu \lesssim$ few kHz plausibly realizable in NS crusts. 

Motivated by $a \sim 1$ in Eq.~(\ref{avalue}) from FRB 121102's bursts, we hereafter adopt $a=1$ in Eq.~(\ref{Efunc}) for the FRB luminosity function for all individual repeaters in this paper. 

For each point in the $P-\dot{P}$ diagram, i.e. a magnetar of a fixed $B$ and $P$, we calculate a relative ``acceptance fraction" defined as the (normalized) fraction of amplitudes $\xi$ drawn from distribution Eq.~(\ref{xipdf}) which satisfy $\Theta$:
\begin{equation}
A_f = \frac{1}{A_{f}^0} \int d \xi \,\frac{d N}{d \xi}  \Theta .
\end{equation}
The normalization $A_{f}^0$ is the asymptotic maximum acceptance fraction in the  $\{P, \dot{P}\}$ space for other parameters fixed and $x \ll 1$ sufficiently small. We display the relative $A_f$ in Figure~\ref{ppdotdiagram} with cyan gradients (with contours of constant $A_f$) -- lower values of  $A_f$ clearly trace the death line, while higher values prefer magnetars with long spin periods. Owing to the constraint Eq.~(\ref{efficiencyconstraint}), extremely large values of $B$ (regardless not plausible in terms of formation and NS stability) are disfavored. The location of asymptotic maximum $A_{f}^0$ depends on $B_{\rm gap}$, the minimum $B$ where pair cascades are viable but is immaterial for the relative prospects between different populations. That is, objects with higher $B$ and $P$ are clearly preferred in Figure~\ref{ppdotdiagram}. For the canonical magnetar population, $A_f \sim 10^{-2}-10^{-3}$ for the parameters adopted in this paper, subject to large (systematic) uncertainties associated with the existence of magnetized crusts at a particular $B$. We do not explicitly exclude objects with spin-down ages $P/\dot{P} \gg 10^{10}$ yr because there exist magnetars, such as the central compact object RCW103 \citep{2016ApJ...828L..13R}, which are beyond this limit owing to unusual formation history. That is, limiting characteristic age may not be an appropriate prior on the space of NS FRB hosts.

\section{Ensemble Luminosity Functions and the Local Fluence Distribution of FRBs }
\label{populationE}

\subsection{Evolved Magnetar Populations}

We aim to estimate the FRB luminosity function for a population of magnetars, selecting for different populations in the $P-\dot{P}$ parameter space via a specified probability density. The choice of the probability distribution selects different subpopulations of potential FRB progenitors, which we weight by the relative acceptance fraction. This allows us to predict normalized luminosity functions for different subpopulations of NSs in the $P-\dot{P}$ diagram, as well as marking nearby objects for further scrutiny. One such choice is the known galactic magnetar population, which occupies a preferential space in $P-\dot{P}$. This choice yields the luminosity distribution of FRBs from a galaxies with similar magnetar production as ours. Since the absolute normalization of the acceptance fraction is poorly constrained, we normalize the simulated luminosity function.

The density of magnetars can be predicted by the rate and evolution history of magnetars, which are then constrained from observations of the galactic population. \cite{2019MNRAS.487.1426B} have established that the population is well described by a model with a decaying surface magnetic field (assuming a uniform birth rate). The main parameters of the model are the range of initial magnetar magnetic fields, $B_0$, the typical time-scale for magnetic field decay, $\tau_B$ and the evolution parameter $\alpha$ which is defined through the relation $\dot{B}\propto B^{1+\alpha}$ \citep{2000ApJ...529L..29C}. While the first parameters are well constrained by observations, \cite{2019MNRAS.487.1426B} found that $\alpha$ can be more prone to observational selection effects,  constrained in the range $-1\leq \alpha\leq 1$. We explore here three characteristic values $\alpha=\{ -1,0,1\}$. We also take $\tau_B=10^4$ yr and $B_0$ to be in the range $B_0=3\times 10^{14}-10^{15}$ G -- these are values generally consistent with the observed magnetar population \citep{2019MNRAS.487.1426B}. As a comparison case explore also an ``extended"  birth distribution $B_0=3\times 10^{14}-3\times10^{15}$ G which may accommodate PSR J0250+5854 \citep{2018ApJ...866...54T} or populations which are broader than that inferred for our galaxy.

The density of magnetars for a selection of these $\alpha$ population models is depicted in Figure~\ref{ppdotdiagram}, in red, blue or green contours along with the known magnetar population in heavy black points. These clearly overlap with the cyan contours of constant acceptance fraction, but with a preference towards higher $P$. The combination of the magnetar population with the acceptance fraction then selects different subpopulations of evolved magnetars which are potential FRB hosts. This subpopulation of preferred FRB hosts is then the probability of finding an evolved magnetar with particular $\{P,\dot{P} \}$ multiplied by the (assumed independent) acceptance fraction $A_f$. Contours for the distribution of this likely FRB host population are depicted in black in Figure~\ref{ppdotdiagram}, along with red/yellow gradients. In all cases, a preference towards objects with higher $P$ is obvious. 
The $dN/dB$ distribution for FRB-mode magnetars, i.e. those which undergo charge-starved short bursts, could be different than the locally inferred evolved magnetar population. We hereafter regard different $dN/dB$ distributions set by $\alpha$ as {\it{not}} strictly interpreted as arising by magnetic evolution, but a convenient parameterization of the FRB host magnetar population.

\subsection{Finite Low Twist}
\label{finitetwist}

Finite low twist (a hierarchy $\rho_{\rm burst} > \rho_{\rm twist} > \rho_{\rm corot}$) will truncate the phase space of allowable amplitudes where charge starvation during bursts will result. Consequently, the death line will be curtailed at some large $P$ if a minimum finite twist is a universal aspect in evolved magnetars. This hierarchy may also be present in certain locales of magnetospheres, e.g. fields connecting to the null charge surface or where the current required to support a twist varies more strongly magnetic colatitude than corotation. Following \citetalias{2019ApJ...879....4W}, the scale at which corotation rather than twist is a limiting factor is,
\begin{equation}
P < P_{\rm twist} \sim \frac{4 \pi R_*}{c \Delta \phi \sin^2 \theta_0} \approx 14 \, \left(\Delta \phi_{-3}\right)^{-1} \quad \rm s
\end{equation}
where we have assumed $\sin \theta_0 \sim 0.2$. For an object such as RCW103, the maximum twist is of order $\Delta \phi \lesssim 10^{-7}-10^{-6}$. Alternatively, for the shortest periods Eq.~(\ref{Pstarve}), the twist is limited to $\Delta \phi \lesssim 2 \pi \nu R_* \sigma_{\rm max}/(c\sin^2 \theta_0) \sim 2 \times 10^{-3}/\sin^2 \theta_0$. 

In our model, the unknown state of finite low twists in the individual magnetars, and the evolution of those finite twists will then impact the total FRB luminosity function over the population of evolved magnetars. The limit of vanishing twist adopted in this paper is then the broadest possible space phase to zeroth order (without invoking spatial variation of the corotation density). 

\subsection{Energy/Luminosity Function}

Now we consider the energy/luminosity function in the elementary model of \S\ref{reverseengineering}. {{We emphasize that the following is only relevant for FRB repeaters.} We sample $\{ P, \dot{P} \}$ values from the probability density of evolved magnetars as determined by the choice of $\alpha$ from the model detailed in  \cite{2019MNRAS.487.1426B}. For each choice of $\{ P, \dot{P} \}$, we compute the individual-repeater energy distribution by sampling Eq.~(\ref{xipdf}) into Eq.~(\ref{Efunc}). Owing to the nontrivial nature of $\Theta$ in Eq.~(\ref{Thetafunc}), a Monte Carlo treatment is expedient. For several million samples, we then obtain a list of energies which we display in Figure~\ref{Edists}. The curves in Figure~\ref{Edists} correspond to the normalized intrinsic FRB luminosity function for a large statistical ensemble of bursts. As is readily apparent, for all choices of $\alpha$, the domain of energies with nonnegligible probability is relatively narrow, about $\sim 1-3$ decades in energy for the full width half maximum. Then, most {{recurrent}} FRBs ought to have energies near the peak of the distributions. We display the approximate\footnote{Assuming a flat spectral index over a respective bandwidth (and fluence range if repeating).} energies of FRBs with apparent localizations \citep[see][]{1980ApJ...236L.109L,2017ApJ...834L...7T,Bannistereaaw5903,2019Natur.572..352R,Prochaska231,Marcote2020} in Figure~\ref{Edists}. {{We omit the low-DM burst FRB 171020 \citep{2018Natur.562..386S} in the Figure~\ref{Edists}, since its localization is less secure than the other bursts. Yet, for the host argued by \cite{2018ApJ...867L..10M}, the estimated isotropic equivalent energy is $\sim 10^{37}-10^{38}$ erg. The model narrowness is largely consistent with most inferred FRB energies, except perhaps for model $\alpha = 1$ if FRB 190523 is regarded as of the same population as repeaters FRB 121102 and FRB 180916}}. Notably, although the ``normal" and ``extended" birth $B_0$ ranges do shift the distributions, the value of $\alpha$ clearly has the greatest impact on the {\it shape and width} of these FRB luminosity functions. {For instance, a preferentially high period and magnetic field sample, as in the model II of $\alpha =-1$, broadens the model luminosity function over that of model I.}

If some subset of FRBs which have not yet repeated, such as FRB 190523, are a different population than considered here, then they may not be as constraining to the luminosity function as suggested by Figure~\ref{Edists}. {{FRB beaming is also allowed in our model (but poorly understood) since the local observed galactic magnetar short burst rate is higher than the observed FRB rate (see \S\ref{FRB190523}).}}

\begin{figure*}
\gridline{\fig{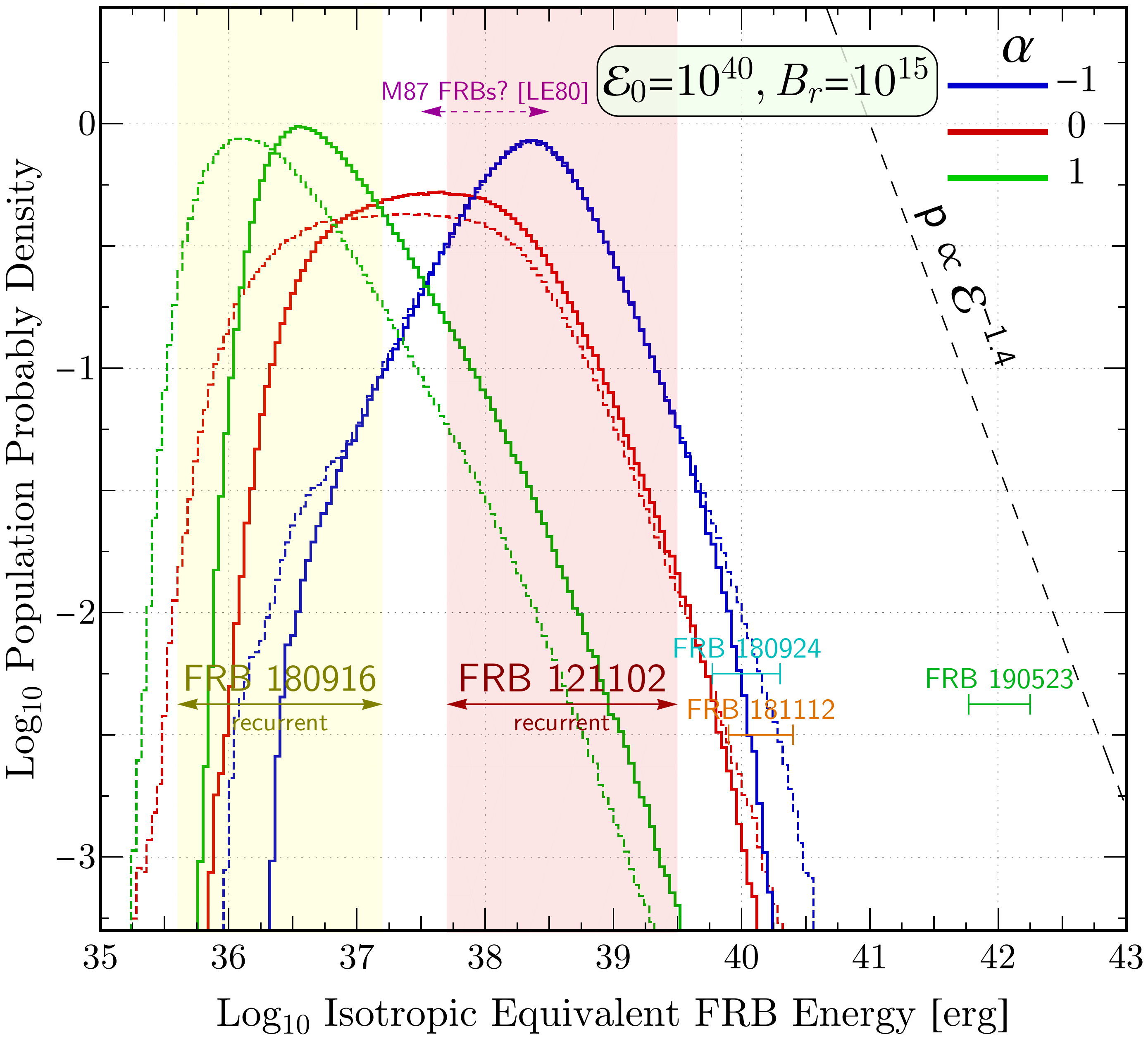}{0.495\textwidth}{}
		\fig{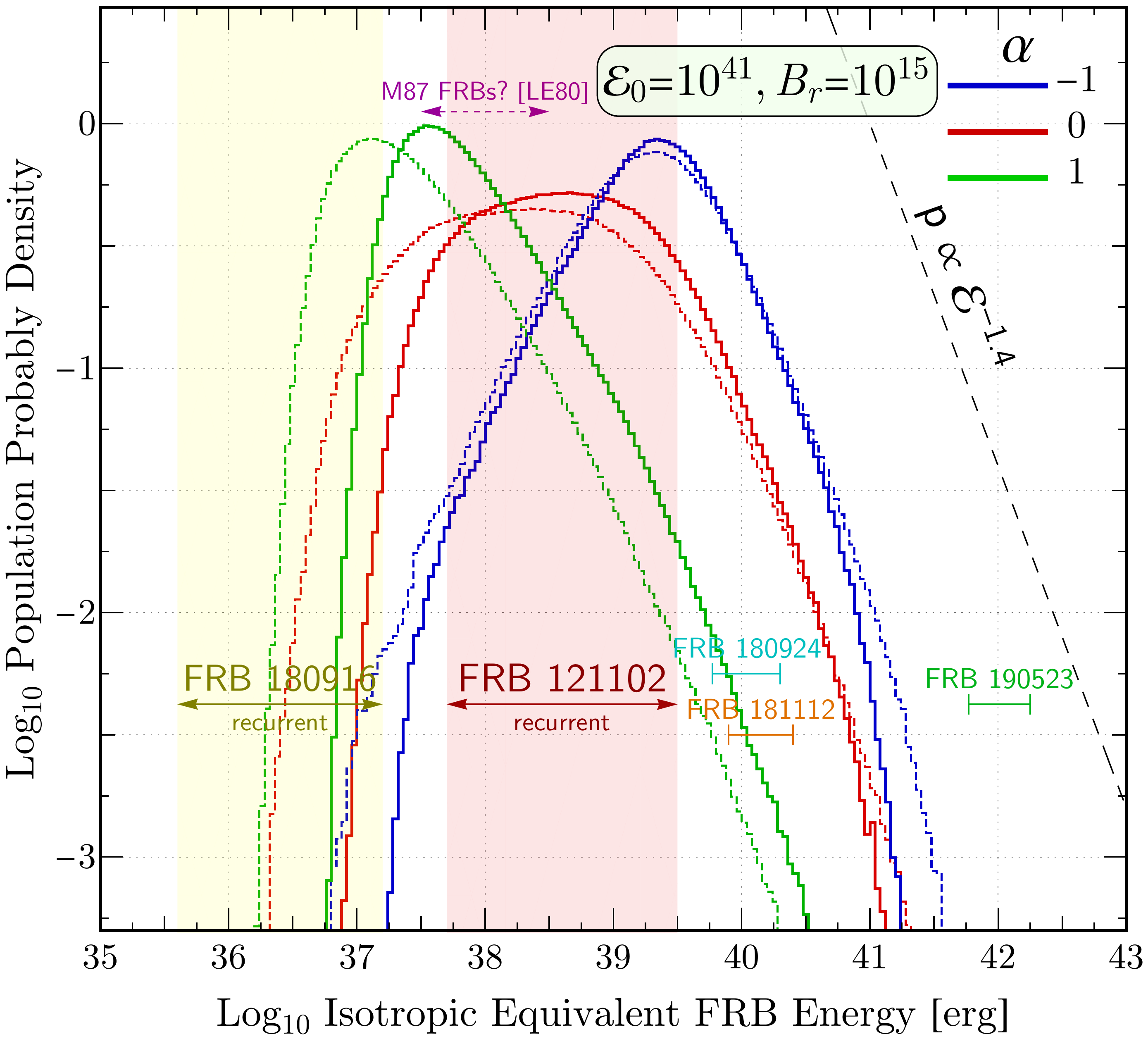}{0.495\textwidth}{}
		}
\gridline{\fig{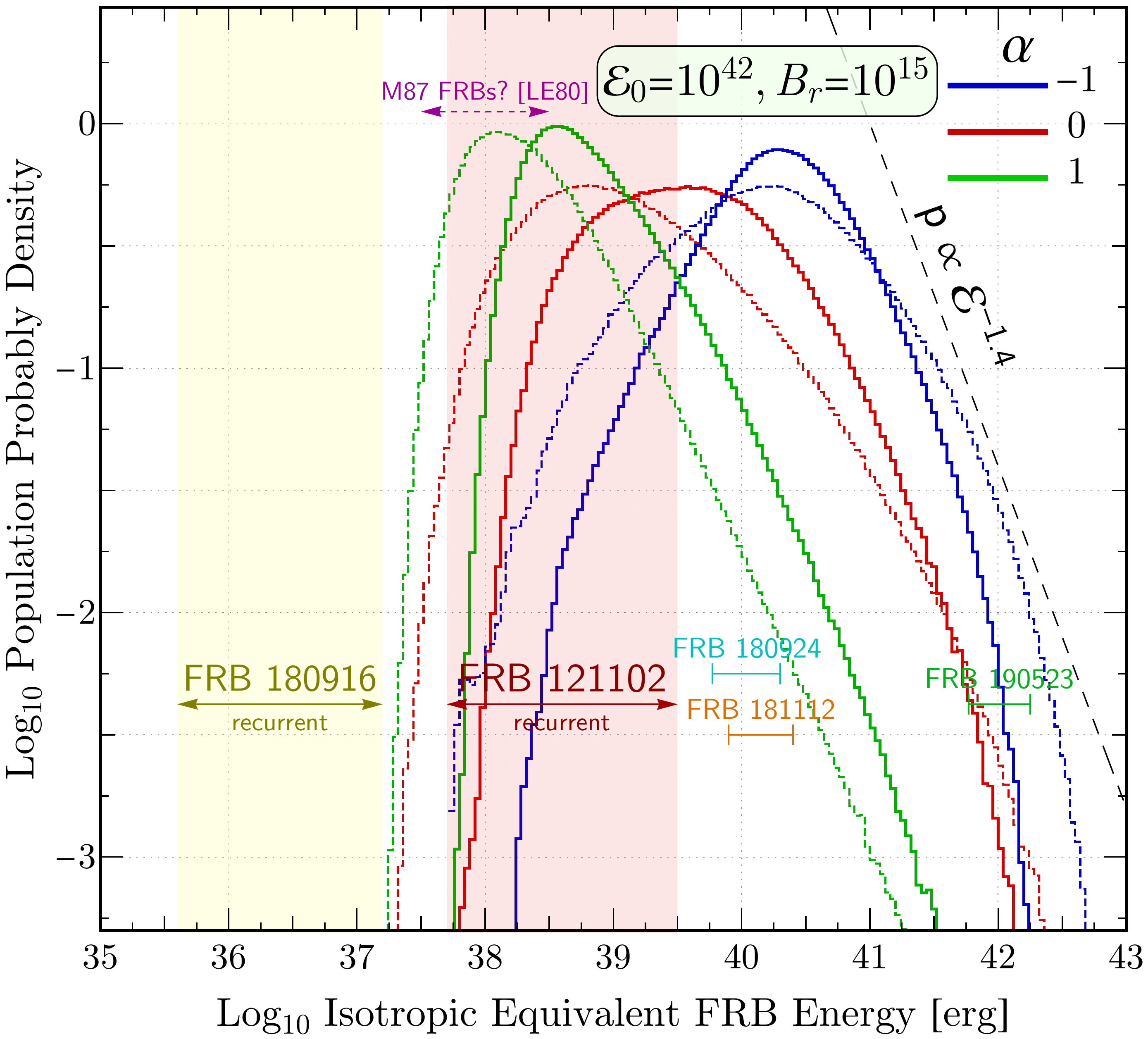}{0.495\textwidth}{}
		}	
\caption{The energy/luminosity function of recurrent FRBs in the low-twist magnetar model, as obtained by convolving the single burst distributions to the model population distributions. The luminosity functions are approximately broken power-law in character, as discussed in the text. The two panels are calculated using different values of the pair $[{\cal E}_0, B_r ]$ as indicated, and other parameters set in Table~\ref{tab:paras}. The solid and dashed lines are models for ``normal" and ``extended" birth $B_0$ values for the population synthesis, respectively. The {crude energy domain of localized recurrent FRBs and approximate energies for other localized but not-yet-recurred bursts are depicted.} As more FRBs are localized in the future, it may be feasible to distinguish among different population models.
\label{Edists}}
\end{figure*}

The shape and limited domain of these energy distributions may be understood as follows. It is evident that the distributions attain power-law regimes -- these arise from the distribution of $B$ in the different $\alpha$ subpopulations. Since $\dot{B}\propto B^{1+\alpha}$, for a constant birth rate we sample $dN/dB \propto B^{-(1+\alpha)}$. For the adopted populations, there are two associated breaks corresponding to low ($B \sim 10^{13}$ G) and high ($B \sim 10^{15}$ G) limits of $dN/dB$ distribution for progenitors corresponding to low and high energy breaks in the FRB energy ${\cal E}$, respectively. That is, when $P \sim 10$ s, the smallest $\xi \sim 10^{-2} \xi_{\rm max}$ which dominate the statistics, the break occurs at ${\cal E}_{\rm high} \sim 10^{-2} {\cal E}_0 (10^{15}/B_r)$ while the lower energy break manifests at  ${\cal E}_{\rm low} \sim 10^{-2} {\cal E}_0 (10^{13}/B_r)$. Below the low-energy break, there is a sharp cut-off owing to the paucity of magnetars at lower $B$. Beyond the high-energy break, the slope corresponds to the adopted amplitude distribution Eq.~(\ref{dndEr}) for the highest $B$ in the sample. That is, above the highest $B \sim 10^{15}$ G, the energy distribution $dN/d\log{\cal E} \propto {\cal E}^{-\Gamma+1}$ steepens to that of individual repeaters' statistics, $ -\Gamma + 1 = -2(s-1)/a \approx -1.4$ for $a=1$ and $s=1.7$. Between the low and high energy breaks, the index corresponds to the adopted $\alpha$ model, $ dN/d{\cal E}_r \propto {\cal E}_r^{-(\alpha+a)/a}$ so that $dN/d\log{\cal E} \propto {\cal E}_r^{-\alpha}$ when $a=1$ -- in this regime, $B$ varies while $\xi$ is roughly constant at the lowest viable value. For instance, a plateau is apparent for the $\alpha =0$ case illustrated in red. For the ``extended" case II, the entire preferred sample shifts to higher $P$ values by about half a decade, as in Figure~\ref{ppdotdiagram} -- this allows for smaller $\xi$ amplitudes to dominate the sample, and correspondingly shifts the peak of the population luminosity function in Figure~\ref{Edists} to lower energies also by about half a decade owing to $\epsilon_1$ in Eq~(\ref{epsilonEcuts}). 

For high FRB energies, the $\Gamma\sim 2.4$ index is somewhat steeper but consistent within uncertainties to that inferred empirically by \cite{2018MNRAS.481.2320L} in their suite of electron models for inferring FRB distances. By construction, \cite{2018MNRAS.481.2320L} did not consider humped or broken power-law luminosity functions. An insufficient number of observed FRBs, which by nature are likely to be observed at the peak (by definition flatter) rather than the tail of the luminosity function, may also imply a flatter inferred luminosity function than reality if a single power-law model is assumed. This motivates inclusion of broken power law models into future luminosity function inferential studies.

\subsection{Local Fluence Distribution of FRBs in Standard Cosmology}
\begin{figure*}
\gridline{\fig{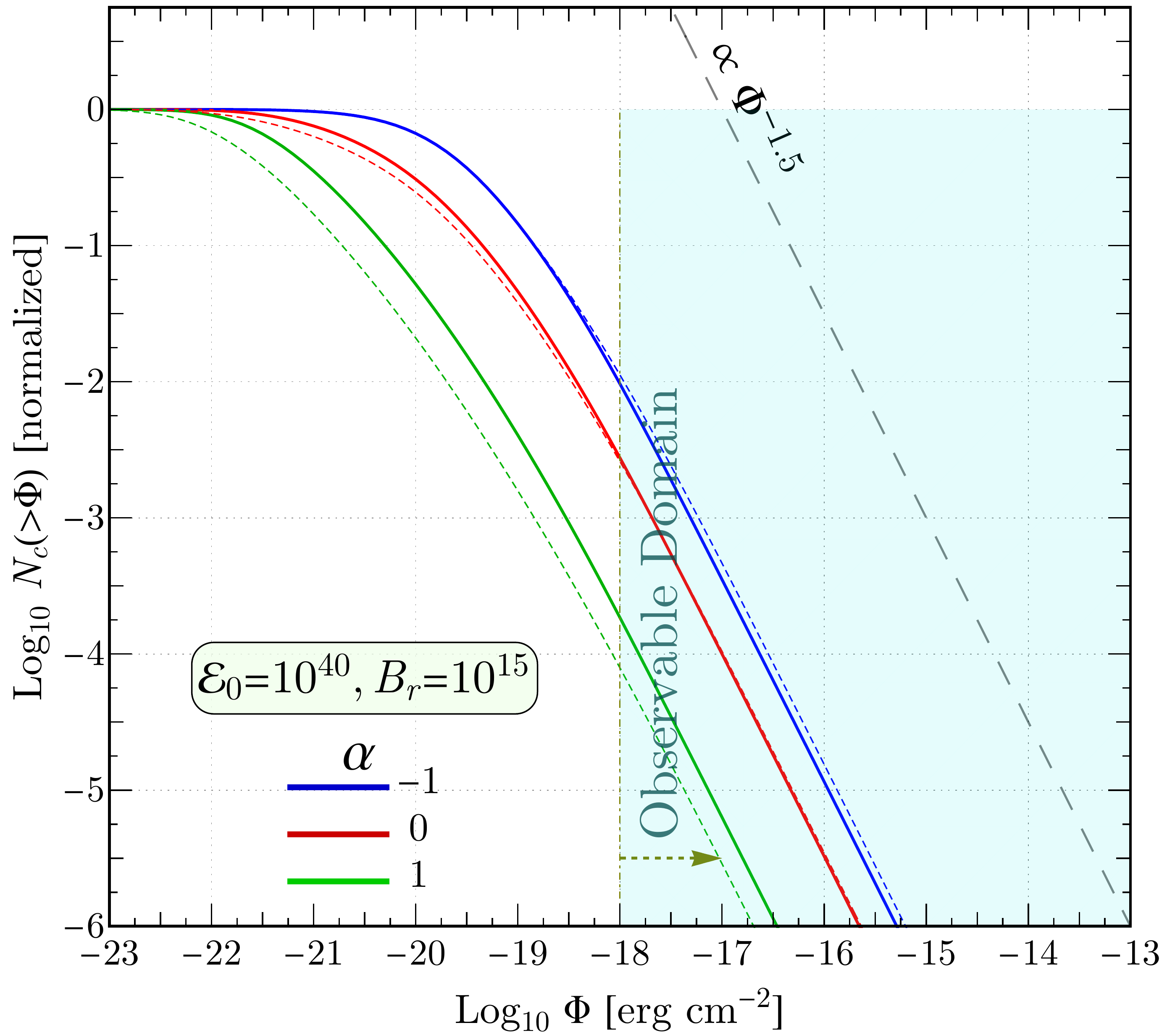}{0.495\textwidth}{}
		\fig{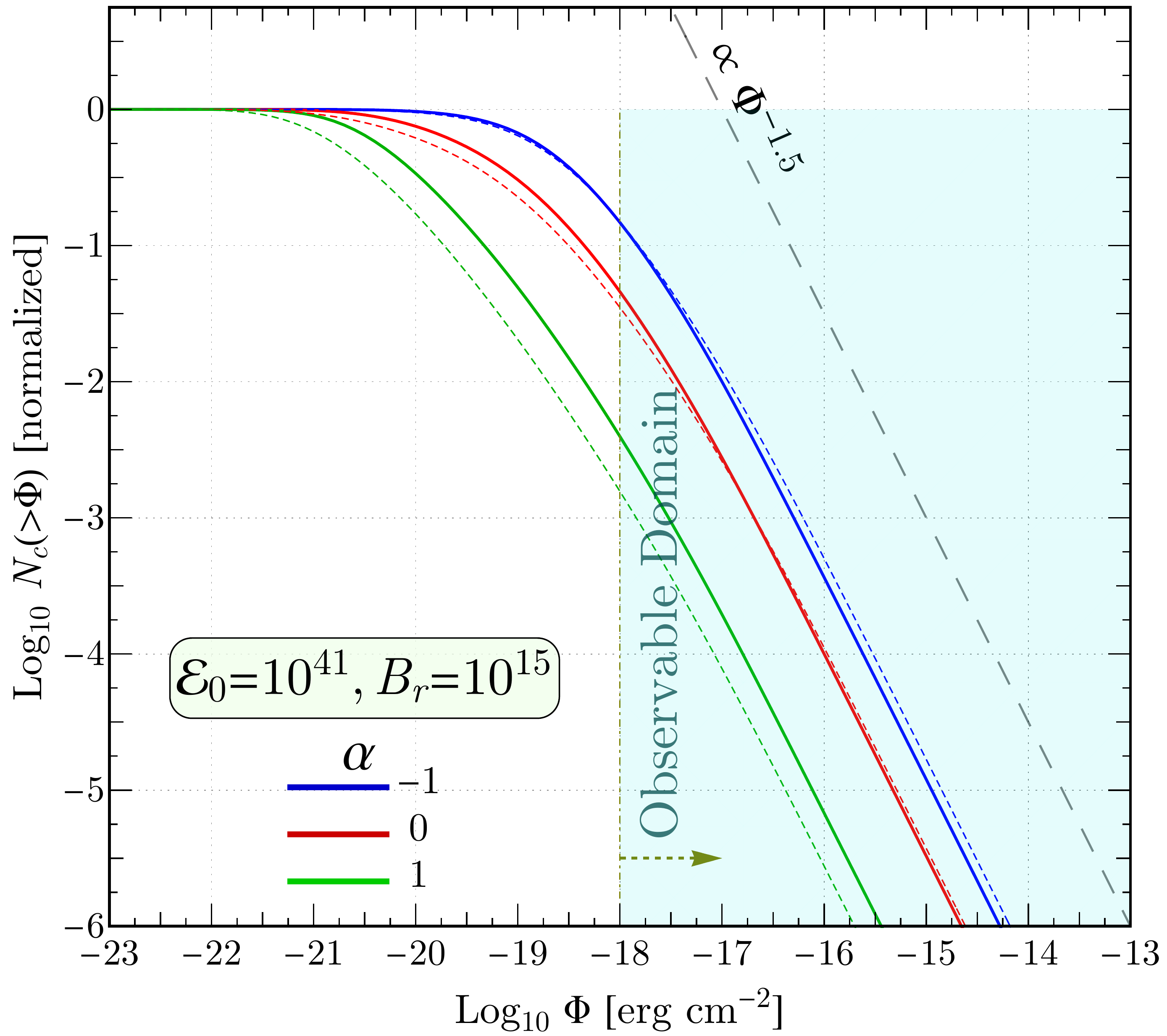}{0.495\textwidth}{}
		}
\gridline{\fig{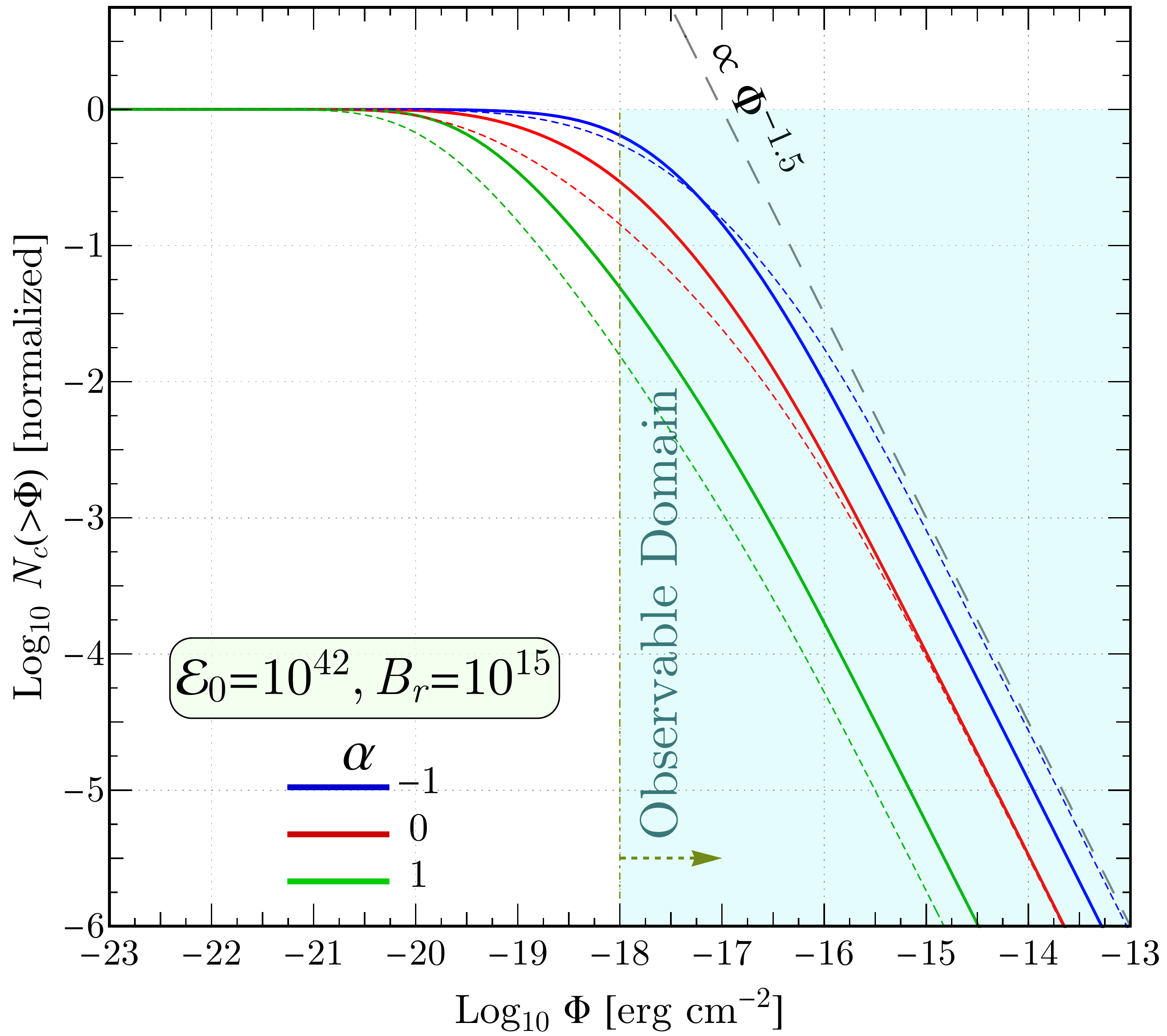}{0.495\textwidth}{}
		}	
\caption{The local fluence-flux distribution from cosmologically distributed FRB-producing magnetars following the star formation rate, sampled from the population FRB luminosity function displayed in Figure~\ref{Edists} -- see text following Eq.~(\ref{Nltp}). The vertical line at $\Phi = 10^{-18}$ erg cm${}^{-2}$ demarcates the typical sensitivity to FRBs for current radio facilities, corresponding to FRB fluence of $\sim 100$ Jy $\mu$s with flat spectral index in a bandwidth of $\nu_r \sim 1$ GHz. The solid and dashed lines are for ``normal" and ``extended" birth $B_0$ values for the population synthesis, respectively. The different panels are calculated using different values of the pair $[{\cal E}_0, B_r ]$ as indicated, and other parameters set in Table~\ref{tab:paras} for individual repeaters. 
\label{localfluencedist}}
\end{figure*}
Let us consider the scenario where {{\it all}} FRBs arise from short bursts in low-twist magnetars. The observed fluence distribution of FRBs is a convolution of the intrinsic rate and the energy distribution function. We take ${\cal N}(<{\cal E})$ as the number of events with an energy of ${\cal E}$ or less and $R_{\rm FRB}(z)$ as the co-moving space density of FRBs in a given redshift interval. That is, ${\cal N}(< {\cal E})$ is the cumulative distribution function of the normalized probability distributions depicted in Figure~\ref{Edists}. The differential co-moving rate of bursts at a given redshift is $R(z)=[ R_{\rm FRB}(z)/(1+z) ] dV(z)/dz $ where $dV(z)/dz$ is the comoving volume element. Then, for luminosity distance $d_L(z)$, the observed fluence cumulative distribution is
\begin{equation}
N_c(>\Phi)\propto \int_0^{\infty} R(z) \left[ 1- {\cal N} \left(<\frac{4\pi d_L(z)^2 \Phi}{1+z} \right) \right]dz.
\label{Nltp}
\end{equation}
We adopt a flat $\Lambda$CDM universe with $\Omega_\Lambda = 0.69$, $\Omega_m = 0.31$ and $H_0 = 68$ km s$^{-1}$ Mpc$^{-1}$ \citep{2016A&A...594A..13P} with $m(z) = \sqrt{\Omega_m (1+z)^3 + \Omega_\Lambda}$ where
\begin{eqnarray}
\frac{dV(z)}{dz}  = \frac{4 \pi c}{H_0} \frac{d_c^2}{m(z)}  \qquad d_c = \frac{c}{H_0} \int_0^z \frac{dz^\prime}{m(z^\prime)}
\end{eqnarray}
and $d_L = (1+z) d_c$. As a first estimate, we consider the case where $R_{\rm FRB}(z)$ follows the star formation rate (SFR) density of \cite{2014ARA&A..52..415M} Eq.~(15), which peaks at $z \sim 2$. The magnetar formation rate, via core collapse supernovae, is a small fraction of the SFR \citep{2019MNRAS.487.1426B}. 

The computation of the distribution Eq.~(\ref{Nltp}) is plotted in Figure~\ref{localfluencedist} for identical parameters as Figure~\ref{Edists}. Owing to the relative narrowness of the luminosity functions in Figure~\ref{Edists}, the standard (``Log N - Log S") $N_c (>\Phi)\propto \Phi^{-1.5}$ is obtained for isotropically-distributed sources (in Euclidean spacetime) with delta function luminosity at high fluences, which correspond to nearby FRBs. For fainter sources, our model predicts a flattening of the index, implying a saturation of FRB rates with improving sensitivity. Observational rate studies \citep[cf. Table 3 in][]{2019A&ARv..27....4P} estimate an all-sky rate of order $\sim 10^{3}-10^4$ FRBs per day above a few Jy ms fluence threshold, corresponding to a level of about $\Phi \sim 10^{-16}-10^{-17}$ erg cm$^{-2}$ for standard assumptions on bandwidth and flat FRB spectral index. As apparent in Figure~\ref{localfluencedist}, we predict that future enhancements in sensitivity may result in up to ${\cal O}(10^2-10^4)$ factor increase in FRB rate over current limits for $\alpha = 1$, while a more modest ${\cal O}(10^1-10^2)$ factor for $\alpha = -1$. Moreover,  for the $\alpha \sim 0  - 1$ cases, deviations from $ d\log N_c (>\Phi)/d\log \Phi > -1.5 $ may become accessible for samples of low fluence bursts. The $\Phi$ dependence of this turnover may also be used to constrain $\alpha$ and other parameters, thereby discriminating among the putative $B$ distributions of low-twist magnetar FRB hosts. 

\newpage
\section{Summary and Discussion}
\label{summarydiscussion}

The low-twist magnetar model \citetalias{2019ApJ...879....4W} hypothesizes a similar trigger for recurrent FRBs as high-energy magnetar short bursts, with the viability of FRB production governed by paucity of charge carriers during field dislocations. In the model, all recurrent FRBs should be accompanied by high-energy quasi-thermal short bursts of energy ${\cal E}_{\rm sb} \lesssim 10^{42}-10^{43}$ erg, but not all short bursts will  produce FRBs, contingent on the state of the magnetar's magnetosphere prior to the burst. In either instance, the energy is predominantly released in the closed zone of the magnetosphere owing to the small polar cap radius. By simple order-of-magnitude scaling relations on parameters known from magnetar short burst phenomenology, we constrain the plausible period and surface magnetic field of FRB host magnetars. Adopting models for evolved magnetar populations, which are constrained from the observed magnetar distribution in our galaxy, we consider the space of likely FRB-mode magnetars in this subset.

Key results include:

\begin{itemize}
\item For other quantities fixed and magnetospheric twist sufficiently small, high $B$ and $P$ are clearly preferred for FRB host magnetars. This is a consequence of the charge starvation condition, as well as requiring magnetic dominance in arcades of field bundles during bursts. A simple formulation of a death line for FRBs suggests $P\gtrsim 2/(\nu \sigma_{\rm max}) \sim 0.2$ s and $B\, P \gtrsim 6 \times 10^{13}$ G s. 

\item An inversion protocol of the power-law fluence distribution of FRB 121102, under the assumption that magnetar short bursts scale quadratically with field dislocation amplitude, suggests that the FRB energy scales close to linearly with amplitude or, physically, the voltage drop for primary particles, a result similar to that inferred for radio emission in most pulsars\footnote{That is, the Poynting flux scales as $\xi^2$ while the conversion into nonthermal coherent radio emission scales close to $\xi$. Analogously, in $\gamma$-ray pulsars, conversion of $\dot{E}$ into nonthermal gamma-rays is also sublinear  \citep[see Fig.~(9) in][]{2013ApJS..208...17A}. } \citep{2002ApJ...568..289A}. Then, the fluence index for FRBs $\Gamma$ ($dN/d{\cal F} \propto {\cal F}^{-\Gamma}$) is related to the familiar index $s \sim 1.7$ for short bursts,  $\Gamma = (2s +a-2)/a \sim 2.4$, when $a\sim 1$. 

\item In an elementary model for the FRB energy, the predicted energy range for individual repeaters is narrow, of order $[2/(\nu P \sigma_{\rm max})]^{-1} \sim 100-1000$ for the ratio of highest to lowest energy bursts for plausible frequencies $\nu \sim 10^2 -10^3$~Hz  of magnetar crusts, for a breaking strain $\sigma_{\rm max} \sim 0.1$ and $P \sim 10$ s. This appears consistent with the distribution of fluences in FRB 121102 {{\citep{2018ApJ...863....2G,2018ApJ...866..149Z,2019ApJ...877L..19G}, FRB 171019 \citep{2019arXiv190810026K} and CHIME repeaters such as FRB 180916 \citep{2019ApJ...885L..24C,Marcote2020}}}.

\item Extending the elementary model for the FRB energy/luminosity for individual repeaters to models of evolved population of magnetars, we find the energy/luminosity function for a population is described by a broken power law. The low energy indices trace the FRB host magnetar population $B$ distribution while the high energy indices probe individual repeater distributions. Then, inferences of the FRB luminosity function can constrain the character of the FRB hosting magnetar population -- broken power-law models are therefore encouraged for future studies.

\item  The population energy/luminosity function is somewhat narrow, with FWHM of $1-3$ decades with a sharp low-energy cut-off around $10^{37}-10^{38}$ erg. The narrowness implies high probability of observing FRB energy corresponding to the peak of the luminosity function, and lends credence to the radio bursts detected by \cite{1980ApJ...236L.109L} as originating from Virgo. Moreover, the narrowness suggests redshifts may be constrained solely by FRB fluence once the luminosity function is better characterized. 

\item Assuming all FRBs follow the SFR with a luminosity function that follows the models for a population of evolved magnetars, the number distribution of highest fluence $\Phi$ bursts scales as $N_c (>\Phi)\propto \Phi^{-1.5}$. At lower fluences, the rate may saturate, depending on the evolved magnetar population under consideration.

\end{itemize}

There are several caveats worth elaborating. The elementary model motivated by the inversion protocol from FRB 121102 should be considered a zeroth-order approximation -- in particular, we have omitted details about any the observing radio frequency $\nu_r$ selection biases, and assumed that the phenomenology may be broadly similar across episodes and sources, as known for magnetar high-energy short bursts. We regard the model as a first step in describing the gross behavior of a large ensemble of bursts over time spans much longer than the spin period of the putative rotator for a large bandwidth (e.g. $\Delta\nu_r \sim 4-8$ GHz as in \cite{2018ApJ...866..149Z}). As noted in \citetalias{2019ApJ...879....4W}, the frequency dependence likely hinges on the altitude of emission, observer perspective (and therefore spin phase) and colatitude of the active region. These nuances provide opportunity to further characterize and test the model in the future.

There are other ingredients, which currently are unconstrained, that may broaden or narrow the luminosity function in Figure~\ref{Edists}. First, if the field twist is nonzero (yet small enough to allow the \citetalias{2019ApJ...879....4W} mechanism, \S\ref{finitetwist}) for some FRB host magnetars, then the range of fluences in those individual repeaters can be significantly narrowed. If that population dominates in number over FRB host magnetars with nearly twist-free magnetospheres, then the population luminosity function will also be narrowed. Second, if a range of eigenmodes and length scales departing from Table~\ref{tab:paras} are realized, particularly at kilohertz frequencies or moderately different active region areas, this may extend the allowed space thereby broadening the luminosity function. The death line in Figure~\ref{ppdotdiagram} would then be a band, rather than a strict demarcation. Third, if beaming is a significant factor, for instance fluence varying with observer magnetic impact angle, then this may broaden the radio energy range of individual repeaters, thereby broadening the population luminosity function. For FRB 121102, the similar arrival time statistics to magnetar short bursts suggests beaming is broad or the geometry special, for instance with a small magnetic obliquity as well as small angle with respect to the spin axis. Fourth, as is already apparent in Figure~\ref{Edists}, broadening the period or B distribution of progenitors will broaden the luminosity function.

Finally, we note that we have assumed the FRB-mode magnetar formation rate follows the core-collapse supernovae rate which is known to conform to the SFR. Two recently localized FRBs \citep[][]{Bannistereaaw5903,2019Natur.572..352R} are hosted by galaxies whose SFRs are significantly lower than typical or that for FRB 121102 \citep{2017ApJ...834L...7T}. If such a trend is sustained with additional FRB localizations, this may suggest a different formation channel for low-twist magnetars with long periods {{\citep[e.g.,][or objects such as in RCW103]{2019arXiv191208287K}}}. 

\subsection{FRB 190523}
\label{FRB190523}

{{FRB 190523 is at the extremity of the model luminosity functions in Figure~\ref{Edists}. In the context of the low-twist model, it could be described as a low (but not hugely so) probability event, arising in the tail of the amplitude distribution in the most extreme magnetars of fields $\sim {\rm few} \times 10^{15}$ G. Of course, hard-to-quantify selection effects may play a role for the localization of FRBs with atypically high energetics, complicating assessments of probability. A beaming factor $f_b<1$ could also be invoked in the model, at the cost of higher cosmological total rates. Such beaming is not issue for the energetics in our model: most bursts would occur well below a cut-off energy scale which dominates the energetics (recall $ d\log N/d\log{\cal E}_{\rm sb} > -2$). Indeed, the local magnetar short burst rate is much higher than the FRB volumetric rate, so some beaming could be tolerated. We are not aware of any study that systematically quantifies the galactic magnetar short burst rate, therefore a limit of the narrowness of beaming cannot currently be quantified. Clearly, a much larger sample of FRBs with precise localizations is necessary to better assess the FRB luminosity function. 

The possibility that repeaters and non-repeaters are different phenomena must be mentioned, including multiple potential populations in either, respectively. Some rate studies \citep{2018ApJ...854L..12P,2019MNRAS.484.5500C,2019arXiv191207847J} seem to indicate that FRB 121102 is perhaps more active than other bursters and exceptional. However, the phenomenology of known galactic magnetars are rife with episodes of prolific activity and slumber. In the context of the magnetar model here, FRB 121102 could simply be in a more active state, with the global FRB rate reflecting the rarity (in space and time) of such states over a magnetar's evolution. A study of CHIME bursts \citep{2019NatAs...3..928R} implies a cosmological FRB rate generally higher than core collapse supernovae, demanding recurrences of FRBs in most compact object models.}}

\subsection{Predictions and Observational Tests of the Low-Twist Magnetar Model}

Many of the aspects here were detailed in \citetalias{2019ApJ...879....4W} but are restated and expanded here for completeness. For the limiting (maximum) twist in \citetalias{2019ApJ...879....4W}, the typical electron plasma frequency is $\omega_e \lesssim 10-20$ GHz at the surface and $\sim 1$ GHz at a few NS radii altitude. This sets the characteristic scale for describing either generation or propagation of coherent radio emission for FRBs.

\paragraph{FRB Arrival Times} 
Since the proposed FRB trigger is identical to one in magnetar short bursts, arrival times of recurrent FRBs should exhibit similarity in statistics if FRBs are {\it not} highly beamed. The correspondence between FRB 121102  and magnetar short bursts in \citetalias{2019ApJ...879....4W} and apparent lack of periodicity supports that beaming is not determinative in FRB 121102. That is, FRBs may be observed from almost any spin phase of the rotation, have a wide emission cone and do not have a ``lighthouse" beam which sweeps past the observer -- this is supported by the fixed PA during pulses  in FRB 121102 \citep{2018Natur.553..182M}. {{ If the situation is similar to magnetar short bursts, when only timing information is employed, significantly larger samples of bursts may be required to establish periodicity \citep[e.g.,][]{2018MNRAS.476.1271E}. }}

As in magnetar short bursts, recurrent FRBs ought to be episodic, with long intervals month/year of inactivity spanning to intense short storms persisting for hours. For these burst storms, based on the phenomenology of magnetar short bursts, the FRB arrival times could be loguniform in character. This gives rise to a humped distribution for waiting times which may be described by a lognormal with mean of $\sim 10^2$ s (in the source frame) of width $\sim1$ dex. Moreover, for sufficiently extended burst storms, a power-law relation between arrival times and next-burst waiting times may be indicated if continuous coverage is maintained as in the August 2017 episode of FRB 121102 (see Fig. 2 in \citetalias{2019ApJ...879....4W} and its magnetar short burst parallel Fig. 10 in \cite{2004ApJ...607..959G}).

\paragraph{Multipulse Substructure and Short-Waiting-Time Repetition Clusters}
In FRB 121102 burst storms, $\sim 5\%$ of recurrences are observed with short $10^{-3}-10^{-2}$ s waiting times \citep{2018ApJ...866..149Z,2019ApJ...877L..19G}. In the low-twist model, such short recurrences may be ascribed to crustal oscillations of similar periods which persist after the initial field dislocation. Blobs of pair plasma could be launched in closed zones with small colatitude (and large maximum altitude). Similarly, multipulse substructure in bursts \citep[e.g.,][]{2019Natur.566..235C} could be attributed to such crustal oscillations. Yet, owing to core-crust coupling \citep[e.g.,][]{2006MNRAS.368L..35L,2014ApJ...793..129H,2019ApJ...871...95M}, trains of these short recurrences should not persist for more than $0.2-2$ s. In standard theory, such oscillations are discrete higher harmonics of a fundamental $n=0, \ell=2$ eigenmode, $\nu_\ell \approx \nu_{0,\ell=2} \sqrt{(\ell-1)(\ell+2)}$ or overtones $(n>0$) modes \citep[e.g., \S12.6 in][]{2008LRR....11...10C}. A spectrum of inferred frequencies in multipulse structure or short-waiting-time-event clusters that match such discrete orderings would be strong evidence of magnetar crust activity in FRBs, and also constrain the NS equation of state.

\paragraph{Frequency Drifts}
Frequency drifts appear to be universally nonpositive in FRBs, suggesting radius-to-frequency mapping of plasma rising in altitude in flux tubes, as in pulsars \citep[e.g.,][]{1978ApJ...222.1006C}. In \citetalias{2019ApJ...879....4W} for either small twists or the corotation density, the local plasma frequency scales as $B_l^{1/2}$ where $B_l$ is the local magnetic field at some altitude - this sets the characteristic scaling of the unknown radio propagation or emission processes. The apparent constancy of $\dot{\nu}_r/\nu_r \sim 10^2$ s$^{-1}$ reported by \cite{2019ApJ...876L..23H} for FRB 121102 is suggestive of a characteristic length scale of $c\nu_r/\dot{\nu}_r \sim 10^8$ cm, when modifications due to relativistic bulk motion are neglected. Similarly, CHIME repeaters \citep{2019ApJ...885L..24C} also exhibit $\dot{\nu}_r/\nu_r \sim 10^1-10^2$ s$^{-1}$. This has been noted also independently by \cite{2019arXiv190807313L} {{\cite[see also][]{2019ApJ...876L..15W}}} and is strongly suggestive of a magnetospheric origin of FRBs (the typical light cylinder size of magnetars is much larger at $c P/(2 \pi) \sim 10^{11}$ cm).

\paragraph{Polarization}
High linear polarization can be generated or imprinted by the strongly magnetized and ordered plasma in NS magnetospheres \citep[e.g.,][]{1979AuJPh..32...61M}. For instance, the X-mode may wave mode couple with plasma on a field loop prior to a decoupling at higher altitudes at the polarization limiting radius \citep{1979ApJ...229..348C}. Small portions of circular polarization, as in some pulsars, may also be feasible by propagation effects depending on the plasma parameters and observer geometry \citep[e.g.,][]{1998Ap&SS.262..379L,2000A&A...355.1168P,2010MNRAS.403..569W}. For objects such as RCW103, vacuum birefringence, sampling the so-called vacuum resonance in a phase space of anomalous dispersion \citep[e.g.][]{2002ApJ...566..373L} in high magnetic fields, may compete with plasma effects in the magnetosphere, even for frequencies as low as $\nu_r \sim 1$ GHz if the field twist is very low. Since in the wave mode coupling regime the PA may trace the ordered field geometry relative to the observer at decoupling, the PA is in principle dependent on spin phase of the rotator. Then, the PA variations in a large sample of bursts can not only be a powerful tool to determine the period of the rotator, but also constrain the observer viewing direction and magnetic obliquity, as for instance in the familiar rotating vector model noted in \citetalias{2019ApJ...879....4W}.

\paragraph{Variable DM}
As in standard pulsar magnetospheres, strongly time variable DM is generally not expected. However, if the FRB host magnetar is in a state of untwisting, then an expanding cavity of low twist develops in equatorial zones \citep{2009ApJ...703.1044B,2017ApJ...844..133C}. This could imprint a small secularly changing DM on the timescale of days/weeks associated with untwisting to quiescence in magnetar outbursts \citep[e.g.,][]{2018MNRAS.474..961C} and also quasi-period DM variations at the spin period. From first principles, such a DM component is difficult to estimate since it depends on the spatial distribution of unknown plasma parameters and 3D wave propagation, but is likely small $< 1$ pc cm$^{-3}$, since this is the typical DM variability in the CHIME repeaters \citep{2019ApJ...885L..24C}.

\paragraph{Other Messengers}
Prompt magnetar short bursts of energies $\sim 10^{36}-10^{42}$ erg ought to be associated with FRBs. Magnetar short bursts are $\sim 10-500$ ms in duration and can be described by two-blackbodies with temperatures of $\sim 1-5$ keV and $\sim 10-50$ keV, with similar flux in both components --  therefore, these would manifest as {\it{unusually~soft~short~GRBs}} or {\it{ultrashort~X-ray~transients}}. Owing to the relatively low energetics, high-energy prompt counterparts to FRBs are not expected unless the host is local within a few Mpc \citep[e.g.,][]{2019ApJ...879...40C}. Therefore, concurrent or associated magnetar short bursts are improbable with current high-energy instruments, except possibly for known Galactic magnetars (\S\ref{localtargets}). If the plasma is indeed confined during bursts in the magnetosphere, no emission beyond a few MeV is expected, owing to high QED opacities in magnetar magnetospheres \citep{2019MNRAS.tmp..982H}. Yet, in the optical/UV the Rayleigh-Jeans tail of the lower temperature short burst component may be detectable as sub-second transients to greater distances with large-sized imaging atmospheric Cherenkov telescopes (or other large optical facilities) in a fast photometry mode \citep[e.g.][]{2009APh....31..156D,2011MNRAS.416.3075L,2019ICRC...36..692H,2019NatAs...3..511B,2018MNRAS.481.2479M,2019ICRC...36..697H}. Finally, it has been long hypothesized that magnetar crustal activity may result in burst gravitational wave (GW) emission \citep{1980Natur.287..122R}. Indeed, a breaking strain of $\sigma_{\rm max} \sim 0.1$ can support large NS ellipticities \citep{2009PhRvL.102s1102H}{{ although searches for burst GW signals concurrent with galactic magnetar short bursts \citep{2011ApJ...734L..35A,2019ApJ...874..163A} argues for null detections of such concurrent GWs for extragalactic FRBs except for extraordinary events.}}

\paragraph{Falsifiability} {{There are several ways the model could be falsified, particularly the simplest variant we have presented in this work. The generally low energy of magnetar short bursts implies no \emph{prompt} X-ray (at energies below about $300$ keV) counterparts  to distant FRBs such as FRB 121102 or FRB 180916 -- any significant detection would immediately falsify the core assumption of the model. Ergo, for distant FRBs, only radio observables are largely accessible in the model. Magnetar giant flares, which can be accompanied by storms of short bursts \citep[e.g.,][]{1999Natur.397...41H} could in be contemporaneous with some nearby recurrent FRB storms in individual repeaters, but would not recur with each FRB pulse. For repeaters, the charge starvation and efficiency constraints Eq.~(\ref{Ecuts})--(\ref{epsilonEcuts}) imply a lower energy scale. The magnetar nature of model also implies an upper scale. Consequently, the range of fluences in individual or a population of repeaters is a key radio observable. If no low energy break in the luminosity function for individual or population of repeaters is found for $\gtrsim 4$ decades in dynamic range of fluence for a sufficiently deep sample would clearly be problematic for the model without invoking beaming or high-period magnetars (e.g. RCW103). If beaming is important, frequency selection effects could be a strong influence (owing to radius-to-frequency mapping) and samples over large bandwidths may be necessary for discrimination. Obviously, nearby FRBs with concurrent messengers offer the most promise to constrain or falsify the model. 
}}

\subsection{Local Targets for Scrutiny}
\label{localtargets}

Given the $\sim 10^3-10^4$ day$^{-1}$ all-sky rate of FRBs, the probability is low that one will be observed locally, given relatively small fields of view of most radio instruments and the $N_c (>\Phi)\propto \Phi^{-1.5}$ scaling. Nevertheless, in our model local magnetars and high-B pulsars may prove interesting for targeted studies -- even modest radio telescopes could be useful owing to the energetics of FRBs. Future observations should scrutinize objects with high $B$ and $P$ simultaneously, especially those that do not exhibit strong nonthermal persistent emission associated with field twists. Transient magnetars\footnote{See the \href{http://www.physics.mcgill.ca/~pulsar/magnetar/main.html}{McGill Magnetar Catalog} \citep{2014ApJS..212....6O} and the \href{https://staff.fnwi.uva.nl/a.l.watts/magnetar/mb.html}{Watts Burst Library}.} of possible interest include 1E 1841--045, 1E 1048.1--5937, SGR 0418+5729, XTE J1810--197 and CXOU J164710.2--455216 among others. Scrutiny of high-B pulsars with exceptionally long periods which lack nonthermal X-ray persistent emission, such as PSR J0250+5854 \citep[$P \approx 23.5$ s and $B \sim 10^{13.5}$ G,][]{2018ApJ...866...54T} or PSR J2251--3711 \citep{2019arXiv191004124M} may also prove interesting. An intriguing possibility is if objects similar to the central compact object RCW103, which has reported magnetar-like short burst activity \citep{2016ApJ...828L..13R} and has an astonishingly large putative spin period of $6.67$ hours, undergo epochs of low twist with impulsive field dislocations. Future radio and high-energy scrutiny of RCW103 and similar objects is also clearly demanded to assess their potential for hosting FRBs.
Finally, given the M87 bursts reported by \cite{1980ApJ...236L.109L}, deep searches of nearby clusters such as Virgo may show success \citep[e.g.,][]{2018ApJ...863..132F} and perhaps offer the prospect of high-energy and optical messengers with existing instruments.

\acknowledgements

We acknowledge helpful discussions with Jonathan Granot, Mansi Kasliwal, Chryssa Kouveliotou, Joeri Van Leeuwen and George Younes. We thank Maxim Lyutikov for alerting us to the reconnection model. Z.W. is supported by the NASA postdoctoral program. A.T. thanks the National Science Foundation for support through grant AST-1616632. M.G.B. thanks the National Science Foundation for support through grant AST-1517550. A.K.H. is supported by the {\it Fermi} Guest Investigator program. This work has made use of the NASA Astrophysics Data System.

\appendix
\section{Constraints by Magnetic Pair Cascade Viability}
\label{pairviability} 

The magnetic field at the surface cannot be too low, otherwise magnetic pair production will not operate efficiently. This obviously is not constraining for magnetars, but may be for older objects. For avalanche magnetic pair production, a conservative estimate is for pair creation above threshold. In magnetar fields, pair creation will occur efficiently at threshold \citep[e.g.,][]{1983ApJ...273..761D,2001ApJ...547..929B}. For above-threshold pair creation in sub-magnetar fields, we require $ \chi = \varepsilon (B/B_{\rm cr}) \sin \theta \gtrsim \chi_0 \approx \max{[0.2, B/B_{\rm cr}]}$ where $\varepsilon \gg 2$ is the photon energy in units of $m_e c^2$, $\theta$ the angle between $B$ and the photon momentum, and $B_{\rm cr} \equiv m_e^2 c^3 /(e\hbar)$ is the quantum critical field. 

The electric field in the gap is established by substitution of $\rho_{\rm burst}$ into Eq.~(27) of \cite{2015ApJ...810..144T}, assuming the gap speed is comparable to $c$:
\begin{equation}
E_{\rm gap} \sim 4 \pi \rho_{\rm burst} \ell_{\rm acc}  \sim \frac{2 \pi \nu B \xi}{c} \frac{\ell_{\rm acc}}{\lambda} \qquad, \qquad \ell_{\rm acc} < \lambda < R_*
\label{Egap}
\end{equation}
where $\ell_{\rm acc}$ is the free acceleration length for a primary electron/positron prior to gap termination, i.e. $d\gamma_e/d\ell_{\rm acc} = -e E_{\rm gap}/(m_e c^2)$, to Lorentz factor,
\begin{equation}
\gamma_{e, \rm acc} = \frac{\pi e \nu B \xi}{m_e c^3 \lambda} \ell_{\rm acc}^2.
\label{gammaacc}
\end{equation}
 
Pair creation by curvature radiation will regulate the acceleration gaps through screening, as in standard pulsar scenarios \citep[e.g.,][]{2015ApJ...810..144T,2019ApJ...871...12T}. 

It can be shown that resonant Compton cooling will not significantly alter particle acceleration before curvature cooling operates for the parameters of interest in this work \citep[see,][]{2011ApJ...733...61B}. 

Curvature radiation reaction is not realized (in the absence of photon splitting, cf. \S\ref{splitting}) before termination of the gap for higher fields ($B \gtrsim 10^{11}$ G) relevant in this work (see Eq.~(54) in \cite{2015ApJ...810..144T}). The characteristic curvature photon energy is $\varepsilon_{\rm CR}  \sim 3 (\lambar/\rho_c)  \gamma_e^3/2$ where $\lambar \equiv \hbar/(m_e c)$ is the reduced Compton wavelength and $\rho_c$ is the curvature radius of an osculating circle for the local field. We note that $\sin \theta \sim \ell_\gamma/\rho_c$ where $\ell_\gamma$ is the characteristic mean free path of magnetic pair creation.

The total gap size is $\ell_{\rm tot} = \ell_{\rm acc} + \ell_\gamma$ minimized for variations in $\ell_{\rm acc}$, as in \cite{2015ApJ...810..144T} and earlier works \citep[e.g.,][]{1998ApJ...508..328H}. Then from the above-threshold pair creation condition $\varepsilon_{\rm CR} (B/B_{\rm cr}) \ell_\gamma/\rho_c \sim \chi_0$, we find
\begin{equation}
\ell_{\gamma, \rm above} \sim \frac{4 \chi_0}{3 \pi^3}\left( \frac{ B_{\rm cr}}{B} \right)^4 \left( \frac{\lambda}{\xi} \right)^3 \left( \frac{\lambar^2 \rho_c^2 c^3}{\nu^3 } \right) \frac{ 1}{\ell_{\rm acc}^6} 
\end{equation}
Minimizing $\ell_{\rm tot}$ for variations in $\ell_{\rm acc}$, we find the total gap size to be 
\begin{eqnarray}
h_{\rm gap, above}  &=&  \frac{7}{3} \left( \frac{2}{\pi} \right)^{3/7} \chi_0^{1/7} \left( \frac{B_{\rm cr}}{B}\right)^{4/7} \left( \frac{\lambda}{\xi} \right)^{3/7} \left(\frac{\rho_c^2 \lambar^2 c^3}{\nu^3}  \right)^{1/7}\\
 & \sim& 2 \times 10^3 \left( \frac{\chi_0 \lambda_{5.5}^3 \rho_{c, 7}^2}{B_{15}^4 \nu_{2}^3 \xi_3^3}  \right)^{1/7} \quad \rm cm
\end{eqnarray}
where we have introduced a factor of two to compensate for relative motion, $h_{\rm gap} \approx 2 \ell_{\rm tot}$.

Adopting the condition $h_{\rm gap} < \delta R_*$ for $\delta R_*/R_* \sim 1$ corresponding to $B$ decreasing by $\sim 1/8$ from the surface to the termination height of the gap, we find a condition on $B$:
\begin{eqnarray}
B \gtrsim B_{\rm gap, above} &=&  \frac{7}{3} \left( \frac{14}{3 \pi} \right)^{3/4} \chi_0^{1/4} \left( \frac{\lambda}{\xi} \right)^{3/4}  \left( \frac{\lambar^2 \rho_c^2 c^3}{\delta R_*^7 \nu^3} \right)^{1/4} B_{\rm cr} \\
&\sim& 1.4 \times 10^{10} \, \, \rho_{c,7}^{1/2} \delta R_{*,5}^{-7/4} \left( \frac{\chi_0 \lambda_{5.5}^3}{\nu_2^3 \xi_3^3} \right)^{1/4} \quad \rm G
\end{eqnarray}
Thus, this bounds the surface field to exceed around $10^{11}$ G. A similar calculation for threshold pair creation $\varepsilon_{\rm CR} \ell_\gamma /\rho_c = 2$ yields,
\begin{eqnarray}
h_{\rm gap, th}  &=& \frac{7}{3} \left( \frac{2}{\pi}\right)^{3/7} \left( \frac{\lambda}{\xi} \right)^{3/7} \left( \frac{B_{\rm cr}}{B} \right)^{3/7} \left( \frac{\lambar^2 \rho_c^2 c^3}{\nu^3} \right)^{1/7} \\
&\sim& 3 \times 10^3 \, \rho_{c,7}^{2/7} \left( \frac{\lambda_{5.5}}{B_{15} \nu_{2} \xi_3} \right)^{3/7} \quad \rm cm
\label{hgapthres}
\end{eqnarray}
with a bound on $B$,
\begin{eqnarray}
B \gtrsim  B_{\rm gap, th} &=& \frac{98}{9 \pi} \left(\frac{7}{3}\right)^{1/3} \frac{\lambda}{\xi} \left( \frac{\lambar^2 \rho_c^2 c^3 }{\delta R_*^7 \nu^3} \right)^{1/3}  B_{\rm cr} \\
&\sim&  10^{9} \, \, \rho_{c,7}^{2/3} \lambda_{5.5} \delta R_{*,5}^{-7/3} \nu_2^{-1} \xi_3^{-1}   \quad \rm G.
\end{eqnarray}
We see that the above-threshold and at-threshold B constraints are similar, and neither demand magnetar-like field strengths for adopted parameters, even for the smaller amplitudes satisfying Eq.~(\ref{ximinP}). We hereafter adopt $B_{\rm gap} = B_{\rm gap, above}$, since threshold pair production is not likely to occur in this lower field regime.

It is interesting to note that for large curvature radii, corresponding to polar locales of field dislocations, the $B_{\rm gap}$ constraints become rather restrictive for small $\xi$. Therefore, the parameter space for FRB producing magnetars is larger for field dislocations away from the polar cap, which may have implications for the PA variations between bursts in individual repeaters.

\section{Influence of Photon Splitting In High Fields}
\label{splitting}

Low-altitude pair cascades may be quenched if all three modes of 
photon splitting allowed by
CP symmetry are operating in the strongly dispersive regime of QED in 
the strong fields
supplied by magnetars. Since photon splitting has no threshold energy, 
it can operate below
the pair creation threshold of $2m_ec^2$.  \cite{2001ApJ...547..929B} 
find that pair cascades
can be strongly suppressed above $B \gtrsim 0.2 B_{\rm cr}$ if all 
three splitting branches are
allowed, provided that $\varepsilon_{\rm CR} \sin \theta < 2$.  What splitting modes operate
in the $B\gg B_{\rm cr}$ domain is currently an open research 
question, with  \cite{1971AnPhy..67..599A}
clearly demonstrating that in weakly dispersive domains, the 
$\perp \to \parallel\parallel$
polarization mode of splitting can proceed in the birefringent, 
magnetized quantum vacuum. An assessment of photon splitting is therefore important for evaluating its potential for influencing pair cascades for field dislocations in the magnetospheres of magnetars.

The length scale $\ell_{\rm sp}$ for splittings can be short, as can be 
discerned from
Fig.~2 of \cite{2019MNRAS.tmp..982H}.  It can be estimated using Eq.~(40) in 
that paper which specialized to a dipole. Adopting
$\theta_{\rm f} \sim R_{\ast}/\rho_c = 0.1 \rho_{c,7}$ for the emission 
colatitude of a curvature
photon of dimensionless  energy $\varepsilon_{\rm CR}$.  Thus, assuming a field value 
of $B = 10^{15}$ G,
\begin{equation}
    \ell_{\rm sp} \;\sim\; 2 \times 10^6 \,  
\frac{\rho_{c,7}^{6/7} }{\varepsilon_{\rm CR}^{5/7}} \qquad
    \; \hbox{cm}.
  \label{eq:ell_sp}
\end{equation}
For splitting to be of relevance, energetic photons $\varepsilon_{\rm 
CR} \gg 2$ ought to split before they have the opportunity to pair produce -- by \S\ref{pairviability}, one anticipates {\it a priori} that $\ell_{\rm sp} \ll 10^4$~cm. If gap formation is inhibited, the electric field Eq.~(\ref{Egap})  will remain unscreened while photon splitting is prolific. The value of $\varepsilon_{\rm CR}$ will be controlled by the competition between electron acceleration and radiative cooling, with the maximum photon energy limited to that in the radiation reaction regime. The characteristic length scale for threshold pair production in a static dipole \citep[the analog of Eq.~(40) in ][] {2019MNRAS.tmp..982H} is approximated by Eqs.~(29)--(30) in \cite{2014ApJ...790...61S}, $\ell_\gamma \sim 32 R_*/(9 \theta_{\rm f} \varepsilon_{\rm CR} [B/B_{\rm cr}]) \propto \varepsilon_{\rm CR}^{-1}$ neglecting the logarithmic dependence. Therefore there exists a value $\varepsilon_{\rm CR} > \varepsilon_{\rm CR, pp}$ such that  $\ell_{\gamma} < \ell_{\rm sp}$ \citep[e.g.,][]{1997ApJ...482..372B}. From Eq.~(\ref{eq:ell_sp}) we obtain
\begin{equation}
 \varepsilon_{\rm CR, pp} \approx 2\times 10^3 \, \rho_{c,7}^{1/2}.
\end{equation}
The value of $\varepsilon_{\rm CR, pp}$ is rather sensitive to the numerical factor in front of Eq.~(\ref{eq:ell_sp}) -- therefore, careful evaluation of the splitting amplitude integrals ${\cal M}_\sigma$ \citep[Eq.~(21) in][]{2019MNRAS.tmp..982H} is essential.

One may adopt 
Eq.~(\ref{Egap}) and equate
the associated acceleration rate $\dot{\gamma}_{\rm acc} \approx e 
E_{\rm gap}/(m_e c)$
to the curvature energy loss rate $\dot{\gamma}_{\rm CR} = (2/3) 
\alpha_f \lambar c \gamma_e^4/\rho_c^2  $,
where $\alpha_f = e^2/\hbar c$ is the fine structure constant. Comparison with Eq.~(\ref{gammaacc}) then yields the characteristic length scale when free acceleration ceases for $B\sim 10^{15}$ G,
\begin{equation}
\ell_{\rm acc, 0} \sim 2 \times 10^{3} \, \lambda_{5.5}^{3/7} \rho_{c,7}^{2/7} B_{15}^{-3/7} \nu_2^{-3/7} \xi_3^{-3/7} \qquad \rm cm.
\label{ellrrla}
\end{equation}
From Eq.~(\ref{gammaacc}), this corresponds to Lorentz factor 
\begin{equation}
\gamma_{\rm e, acc, 0}\sim 8\times 10^7 \, \rho_{c,7}^{4/7} B_{15}^{1/7} \nu_2^{1/7} \xi_3^{1/7} \lambda_{5.5}^{-1/7}
\end{equation}
and curvature radiation photon energy $\varepsilon_{\rm CR}  \sim 3 (\lambar/\rho_c)  \gamma_e^3/2$ 
\begin{equation}
\varepsilon_{\rm CR,0} \sim 3\times 10^6 \, \rho_{c,7}^{5/7} B_{15}^{3/7} \nu_2^{3/7} \xi_3^{3/7} \lambda_{5.5}^{-3/7}
\end{equation}
i.e. with a characteristic energy scale of a TeV. The corresponding pair production length is then $\ell_\gamma \ll 10$ cm for $\rho_c \sim 10^7$ cm. Observe that $\varepsilon_{\rm CR,0} \gg \varepsilon_{\rm CR,pp}$ so that pair production will be immediate, and the splitting length scale significantly larger than $\ell_\gamma$. By {{\it reductio ad absurdum}}, we conclude the primary particle will never enter the regime of radiation reaction before pair production quenches the gap and that splitting will not strongly influence the gap physics for $\rho_c \sim 10^7$~cm. This assessment is perhaps related to the existence of radio-loud magnetars \citep[e.g.,][]{2006Natur.442..892C}, albeit ephemeral radio emitters and suggests that even if all three splitting modes operate, photon splitting likely does not totally quench pair cascades and a pulsar-like mechanism of the proposed FRB model. Yet, splitting may be important in certain parameter regimes for where radio emission or FRBs in magnetars are generated -- polar locales corresponding to larger curvature radii may be impacted by the competition between splitting and pair production, since $\ell_\gamma/\ell_{\rm sp} \propto \rho_c^{1/7}$ and  $\varepsilon_{\rm CR, pp} \propto \rho_c^{1/2}$. There also exist regimes between $B \sim 10^{13} -10^{14}$ G where the splitting length may be shorter than the pair length, owing to the dependence of splitting amplitude integrals ${\cal M}_\sigma$ on $B$ \citep[for details, see][]{2019MNRAS.tmp..982H}. A more complete study of pair cascades with splitting for FRB models is deferred to the future.

\bibliographystyle{aasjournal}
\bibliography{magnetarrefs}

\begin{thebibliography}{}
\expandafter\ifx\csname natexlab\endcsname\relax\def\natexlab#1{#1}\fi
\providecommand{\url}[1]{\href{#1}{#1}}
\providecommand{\dodoi}[1]{doi:~\href{http://doi.org/#1}{\nolinkurl{#1}}}
\providecommand{\doeprint}[1]{\href{http://ascl.net/#1}{\nolinkurl{http://ascl.net/#1}}}
\providecommand{\doarXiv}[1]{\href{https://arxiv.org/abs/#1}{\nolinkurl{https://arxiv.org/abs/#1}}}

\bibitem[{{Abadie} {et~al.}(2011){Abadie}, {Abbott}, {Abbott}, {Abernathy},
  {Accadia}, {Acernese}, {Adams}, \& et~al.}]{2011ApJ...734L..35A}
{Abadie}, J., {Abbott}, B.~P., {Abbott}, R., {et~al.} 2011, \apjl, 734, L35,
  \dodoi{10.1088/2041-8205/734/2/L35}

\bibitem[{{Abbott} {et~al.}(2019){Abbott}, {Abbott}, {Abbott}, {Abraham},
  {Acernese}, {Ackley}, {Adams}, \& et~al.}]{2019ApJ...874..163A}
{Abbott}, B.~P., {Abbott}, R., {Abbott}, T.~D., {et~al.} 2019, \apj, 874, 163,
  \dodoi{10.3847/1538-4357/ab0e15}

\bibitem[{{Abdo} {et~al.}(2013){Abdo}, {Ajello}, {Allafort}, {Baldini},
  {Ballet}, {Barbiellini}, {Baring}, {Bastieri}, {Belfiore}, {Bellazzini},
  {Bhattacharyya}, {Bissaldi}, {Bloom}, {Bonamente}, {Bottacini}, {Brandt},
  {Bregeon}, {Brigida}, {Bruel}, {Buehler}, {Burgay}, {Burnett}, {Busetto},
  {Buson}, {Caliandro}, {Cameron}, {Camilo}, {Caraveo}, {Casandjian}, {Cecchi},
  {{\c{C}}elik}, {Charles}, {Chaty}, {Chaves}, {Chekhtman}, {Chen}, {Chiang},
  {Chiaro}, {Ciprini}, {Claus}, {Cognard}, {Cohen-Tanugi}, {Cominsky},
  {Conrad}, {Cutini}, {D'Ammando}, {de Angelis}, {DeCesar}, {De Luca}, {den
  Hartog}, {de Palma}, {Dermer}, {Desvignes}, {Digel}, {Di Venere}, {Drell},
  {Drlica-Wagner}, {Dubois}, {Dumora}, {Espinoza}, {Falletti}, {Favuzzi},
  {Ferrara}, {Focke}, {Franckowiak}, {Freire}, {Funk}, {Fusco}, {Gargano},
  {Gasparrini}, {Germani}, {Giglietto}, {Giommi}, {Giordano}, {Giroletti},
  {Glanzman}, {Godfrey}, {Gotthelf}, {Grenier}, {Grondin}, {Grove},
  {Guillemot}, {Guiriec}, {Hadasch}, {Hanabata}, {Harding}, {Hayashida},
  {Hays}, {Hessels}, {Hewitt}, {Hill}, {Horan}, {Hou}, {Hughes}, {Jackson},
  {Janssen}, {Jogler}, {J{\'o}hannesson}, {Johnson}, {Johnson}, {Johnson},
  {Johnson}, {Johnston}, {Kamae}, {Kataoka}, {Keith}, {Kerr}, {Kn{\"o}dlseder},
  {Kramer}, {Kuss}, {Lande}, {Larsson}, {Latronico}, {Lemoine-Goumard},
  {Longo}, {Loparco}, {Lovellette}, {Lubrano}, {Lyne}, {Manchester}, {Marelli},
  {Massaro}, {Mayer}, {Mazziotta}, {McEnery}, {McLaughlin}, {Mehault},
  {Michelson}, {Mignani}, {Mitthumsiri}, {Mizuno}, {Moiseev}, {Monzani},
  {Morselli}, {Moskalenko}, {Murgia}, {Nakamori}, {Nemmen}, {Nuss}, {Ohno},
  {Ohsugi}, {Orienti}, {Orlando}, {Ormes}, {Paneque}, {Panetta}, {Parent},
  {Perkins}, {Pesce-Rollins}, {Pierbattista}, {Piron}, {Pivato}, {Pletsch},
  {Porter}, {Possenti}, {Rain{\`o}}, {Rando}, {Ransom}, {Ray}, {Razzano},
  {Rea}, {Reimer}, {Reimer}, {Renault}, {Reposeur}, {Ritz}, {Romani}, {Roth},
  {Rousseau}, {Roy}, {Ruan}, {Sartori}, {Saz Parkinson}, {Scargle}, {Schulz},
  {Sgr{\`o}}, {Shannon}, {Siskind}, {Smith}, {Spandre}, {Spinelli}, {Stappers},
  {Strong}, {Suson}, {Takahashi}, {Thayer}, {Thayer}, {Theureau}, {Thompson},
  {Thorsett}, {Tibaldo}, {Tibolla}, {Tinivella}, {Torres}, {Tosti}, {Troja},
  {Uchiyama}, {Usher}, {Vandenbroucke}, {Vasileiou}, {Venter}, {Vianello},
  {Vitale}, {Wang}, {Weltevrede}, {Winer}, {Wolff}, {Wood}, {Wood}, {Wood}, \&
  {Yang}}]{2013ApJS..208...17A}
{Abdo}, A.~A., {Ajello}, M., {Allafort}, A., {et~al.} 2013, \apjs, 208, 17,
  \dodoi{10.1088/0067-0049/208/2/17}

\bibitem[{{Adler}(1971)}]{1971AnPhy..67..599A}
{Adler}, S.~L. 1971, Annals of Physics, 67, 599,
  \dodoi{10.1016/0003-4916(71)90154-0}

\bibitem[{{Aptekar} {et~al.}(2001){Aptekar}, {Frederiks}, {Golenetskii},
  {Il'inskii}, {Mazets}, {Pal'shin}, {Butterworth}, \&
  {Cline}}]{2001ApJS..137..227A}
{Aptekar}, R.~L., {Frederiks}, D.~D., {Golenetskii}, S.~V., {et~al.} 2001,
  \apjs, 137, 227, \dodoi{10.1086/322530}

\bibitem[{{Arzoumanian} {et~al.}(2002){Arzoumanian}, {Chernoff}, \&
  {Cordes}}]{2002ApJ...568..289A}
{Arzoumanian}, Z., {Chernoff}, D.~F., \& {Cordes}, J.~M. 2002, \apj, 568, 289,
  \dodoi{10.1086/338805}

\bibitem[{Bannister {et~al.}(2019)Bannister, Deller, Phillips, Macquart,
  Prochaska, Tejos, Ryder, Sadler, Shannon, Simha, Day, McQuinn, North-Hickey,
  Bhandari, Arcus, Bennert, Burchett, Bouwhuis, Dodson, Ekers, Farah, Flynn,
  James, Kerr, Lenc, Mahony, O{\textquoteright}Meara, Os{\l}owski, Qiu, Treu,
  U, Bateman, Bock, Bolton, Brown, Bunton, Chippendale, Cooray, Cornwell,
  Gupta, Hayman, Kesteven, Koribalski, MacLeod, McClure-Griffiths, Neuhold,
  Norris, Pilawa, Qiao, Reynolds, Roxby, Shimwell, Voronkov, \&
  Wilson}]{Bannistereaaw5903}
Bannister, K.~W., Deller, A.~T., Phillips, C., {et~al.} 2019, Science,
  \dodoi{10.1126/science.aaw5903}

\bibitem[{{Baring} \& {Harding}(1997)}]{1997ApJ...482..372B}
{Baring}, M.~G., \& {Harding}, A.~K. 1997, \apj, 482, 372,
  \dodoi{10.1086/304152}

\bibitem[{{Baring} \& {Harding}(2001)}]{2001ApJ...547..929B}
---. 2001, \apj, 547, 929, \dodoi{10.1086/318390}

\bibitem[{{Baring} {et~al.}(2011){Baring}, {Wadiasingh}, \&
  {Gonthier}}]{2011ApJ...733...61B}
{Baring}, M.~G., {Wadiasingh}, Z., \& {Gonthier}, P.~L. 2011, \apj, 733, 61,
  \dodoi{10.1088/0004-637X/733/1/61}

\bibitem[{{Beloborodov}(2009)}]{2009ApJ...703.1044B}
{Beloborodov}, A.~M. 2009, \apj, 703, 1044,
  \dodoi{10.1088/0004-637X/703/1/1044}

\bibitem[{{Beloborodov}(2017)}]{2017ApJ...843L..26B}
---. 2017, \apjl, 843, L26, \dodoi{10.3847/2041-8213/aa78f3}

\bibitem[{{Beloborodov} \& {Thompson}(2007)}]{2007ApJ...657..967B}
{Beloborodov}, A.~M., \& {Thompson}, C. 2007, \apj, 657, 967,
  \dodoi{10.1086/508917}

\bibitem[{{Benbow} {et~al.}(2019){Benbow}, {Bird}, {Brill}, {Brose}, {Chromey},
  {Daniel}, {Feng}, {Finley}, {Fortson}, {Furniss}, {Gilland ers}, {Giuri},
  {Gueta}, {Hanna}, {Halpern}, {Hassan}, {Holder}, {Hughes}, {Humensky},
  {Joyce}, {Kaaret}, {Kar}, {Kelley-Hoskins}, {Kertzman}, {Kieda}, {Krause},
  {Lang}, {Lin}, {Maier}, {Matthews}, {Moriarty}, {Mukherjee}, {Nieto},
  {Nievas-Rosillo}, {O'Brien}, {Ong}, {Park}, {Petrashyk}, {Pohl}, {Pueschel},
  {Quinn}, {Ragan}, {Reynolds}, {Richards}, {Roache}, {Rulten}, {Sadeh},
  {Santander}, {Sembroski}, {Shahinyan}, {Sushch}, {Wakely}, {Wells}, {Wilcox},
  {Wilhelm}, {Williams}, \& {Williamson}}]{2019NatAs...3..511B}
{Benbow}, W., {Bird}, R., {Brill}, A., {et~al.} 2019, Nature Astronomy, 3, 511,
  \dodoi{10.1038/s41550-019-0741-z}

\bibitem[{{Beniamini} {et~al.}(2019){Beniamini}, {Hotokezaka}, {van der Horst},
  \& {Kouveliotou}}]{2019MNRAS.487.1426B}
{Beniamini}, P., {Hotokezaka}, K., {van der Horst}, A., \& {Kouveliotou}, C.
  2019, \mnras, 487, 1426, \dodoi{10.1093/mnras/stz1391}

\bibitem[{{Caleb} {et~al.}(2019){Caleb}, {Stappers}, {Rajwade}, \&
  {Flynn}}]{2019MNRAS.484.5500C}
{Caleb}, M., {Stappers}, B.~W., {Rajwade}, K., \& {Flynn}, C. 2019, \mnras,
  484, 5500, \dodoi{10.1093/mnras/stz386}

\bibitem[{{Camilo} {et~al.}(2006){Camilo}, {Ransom}, {Halpern}, {Reynolds},
  {Helfand}, {Zimmerman}, \& {Sarkissian}}]{2006Natur.442..892C}
{Camilo}, F., {Ransom}, S.~M., {Halpern}, J.~P., {et~al.} 2006, \nat, 442, 892,
  \dodoi{10.1038/nature04986}

\bibitem[{{Chamel} \& {Haensel}(2008)}]{2008LRR....11...10C}
{Chamel}, N., \& {Haensel}, P. 2008, Living Reviews in Relativity, 11, 10,
  \dodoi{10.12942/lrr-2008-10}

\bibitem[{{Chen} \& {Beloborodov}(2017)}]{2017ApJ...844..133C}
{Chen}, A.~Y., \& {Beloborodov}, A.~M. 2017, \apj, 844, 133,
  \dodoi{10.3847/1538-4357/aa7a57}

\bibitem[{{Cheng} \& {Ruderman}(1979)}]{1979ApJ...229..348C}
{Cheng}, A.~F., \& {Ruderman}, M.~A. 1979, \apj, 229, 348,
  \dodoi{10.1086/156959}

\bibitem[{{Cheng} {et~al.}(1996){Cheng}, {Epstein}, {Guyer}, \&
  {Young}}]{1996Natur.382..518C}
{Cheng}, B., {Epstein}, R.~I., {Guyer}, R.~A., \& {Young}, A.~C. 1996, \nat,
  382, 518, \dodoi{10.1038/382518a0}

\bibitem[{{CHIME/FRB Collaboration} {et~al.}(2019{\natexlab{a}}){CHIME/FRB
  Collaboration}, {Amiri}, {Bandura}, {Bhardwaj}, {Boubel}, {Boyce}, {Boyle},
  {.~Brar}, {Burhanpurkar}, {Cassanelli}, {Chawla}, {Cliche}, {Cubranic},
  {Deng}, {Denman}, {Dobbs}, {Fandino}, {Fonseca}, {Gaensler}, {Gilbert},
  {Gill}, {Giri}, {Good}, {Halpern}, {Hanna}, {Hill}, {Hinshaw}, {H{\"o}fer},
  {Josephy}, {Kaspi}, {Landecker}, {Lang}, {Lin}, {Masui}, {Mckinven},
  {Mena-Parra}, {Merryfield}, {Michilli}, {Milutinovic}, {Moatti}, {Naidu},
  {Newburgh}, {Ng}, {Patel}, {Pen}, {Pinsonneault-Marotte}, {Pleunis},
  {Rafiei-Ravandi}, {Rahman}, {Ransom}, {Renard}, {Scholz}, {Shaw}, {Siegel},
  {Smith}, {Stairs}, {Tendulkar}, {Tretyakov}, {Vanderlinde}, \&
  {Yadav}}]{2019Natur.566..235C}
{CHIME/FRB Collaboration}, {Amiri}, M., {Bandura}, K., {et~al.}
  2019{\natexlab{a}}, \nat, 566, 235, \dodoi{10.1038/s41586-018-0864-x}

\bibitem[{{CHIME/FRB Collaboration} {et~al.}(2019{\natexlab{b}}){CHIME/FRB
  Collaboration}, {Andersen}, {Bandura}, {Bhardwaj}, {Boubel}, {Boyce},
  {Boyle}, {Brar}, {Cassanelli}, {Chawla}, {Cubranic}, {Deng}, {Dobbs},
  {Fandino}, {Fonseca}, {Gaensler}, {Gilbert}, {Giri}, {Good}, {Halpern},
  {Hill}, {Hinshaw}, {H{\"o}fer}, {Josephy}, {Kaspi}, {Kothes}, {Landecker},
  {Lang}, {Li}, {Lin}, {Masui}, {Mena-Parra}, {Merryfield}, {Mckinven},
  {Michilli}, {Milutinovic}, {Naidu}, {Newburgh}, {Ng}, {Patel}, {Pen},
  {Pinsonneault-Marotte}, {Pleunis}, {Rafiei-Ravandi}, {Rahman}, {Ransom},
  {Renard}, {Scholz}, {Siegel}, {Singh}, {Smith}, {Stairs}, {Tendulkar},
  {Tretyakov}, {Vanderlinde}, {Yadav}, \& {Zwaniga}}]{2019ApJ...885L..24C}
{CHIME/FRB Collaboration}, {Andersen}, B.~C., {Bandura}, K., {et~al.}
  2019{\natexlab{b}}, \apjl, 885, L24, \dodoi{10.3847/2041-8213/ab4a80}

\bibitem[{{Collazzi} {et~al.}(2015){Collazzi}, {Kouveliotou}, {van der Horst},
  {Younes}, {Kaneko}, {G{\"o}{\u g}{\"u}{\c s}}, {Lin}, {Granot}, {Finger},
  {Chaplin}, {Huppenkothen}, {Watts}, {von Kienlin}, {Baring}, {Gruber},
  {Bhat}, {Gibby}, {Gehrels}, {McEnery}, {van der Klis}, \&
  {Wijers}}]{2015ApJS..218...11C}
{Collazzi}, A.~C., {Kouveliotou}, C., {van der Horst}, A.~J., {et~al.} 2015,
  \apjs, 218, 11, \dodoi{10.1088/0067-0049/218/1/11}

\bibitem[{{Colpi} {et~al.}(2000){Colpi}, {Geppert}, \&
  {Page}}]{2000ApJ...529L..29C}
{Colpi}, M., {Geppert}, U., \& {Page}, D. 2000, \apjl, 529, L29,
  \dodoi{10.1086/312448}

\bibitem[{{Cordes}(1978)}]{1978ApJ...222.1006C}
{Cordes}, J.~M. 1978, \apj, 222, 1006, \dodoi{10.1086/156218}

\bibitem[{{Coti Zelati} {et~al.}(2018){Coti Zelati}, {Rea}, {Pons}, {Campana},
  \& {Esposito}}]{2018MNRAS.474..961C}
{Coti Zelati}, F., {Rea}, N., {Pons}, J.~A., {Campana}, S., \& {Esposito}, P.
  2018, \mnras, 474, 961, \dodoi{10.1093/mnras/stx2679}

\bibitem[{{Cunningham} {et~al.}(2019){Cunningham}, {Cenko}, {Burns},
  {Goldstein}, {Lien}, {Kocevski}, {Briggs}, {Connaughton}, {Miller},
  {Racusin}, \& {Stanbro}}]{2019ApJ...879...40C}
{Cunningham}, V., {Cenko}, S.~B., {Burns}, E., {et~al.} 2019, \apj, 879, 40,
  \dodoi{10.3847/1538-4357/ab2235}

\bibitem[{{Daugherty} \& {Harding}(1983)}]{1983ApJ...273..761D}
{Daugherty}, J.~K., \& {Harding}, A.~K. 1983, \apj, 273, 761,
  \dodoi{10.1086/161411}

\bibitem[{{Deil} {et~al.}(2009){Deil}, {Domainko}, {Hermann}, {Clapson},
  {F{\"o}rster}, {van Eldik}, \& {Hofmann}}]{2009APh....31..156D}
{Deil}, C., {Domainko}, W., {Hermann}, G., {et~al.} 2009, Astroparticle
  Physics, 31, 156, \dodoi{10.1016/j.astropartphys.2008.12.008}

\bibitem[{{Elenbaas} {et~al.}(2018){Elenbaas}, {Watts}, \&
  {Huppenkothen}}]{2018MNRAS.476.1271E}
{Elenbaas}, C., {Watts}, A.~L., \& {Huppenkothen}, D. 2018, \mnras, 476, 1271,
  \dodoi{10.1093/mnras/sty321}

\bibitem[{{Fialkov} {et~al.}(2018){Fialkov}, {Loeb}, \&
  {Lorimer}}]{2018ApJ...863..132F}
{Fialkov}, A., {Loeb}, A., \& {Lorimer}, D.~R. 2018, \apj, 863, 132,
  \dodoi{10.3847/1538-4357/aad196}

\bibitem[{{Fonseca} {et~al.}(2020){Fonseca}, {Andersen}, {Bhardwaj}, {Chawla},
  {Good}, {Josephy}, {Kaspi}, {Masui}, {Mckinven}, {Michilli}, {Pleunis},
  {Shin}, {Tendulkar}, {Bandura}, {Boyle}, {Brar}, {Cassanelli}, {Cubranic},
  {Dobbs}, {Dong}, {Gaensler}, {Hinshaw}, {Land ecker}, {Leung}, {Li}, {Lin},
  {Mena-Parra}, {Merryfield}, {Naidu}, {Ng}, {Patel}, {Pen}, {Rafiei-Ravandi},
  {Rahman}, {Ransom}, {Scholz}, {Smith}, {Stairs}, {Vanderlinde}, {Yadav}, \&
  {Zwaniga}}]{2020arXiv200103595F}
{Fonseca}, E., {Andersen}, B.~C., {Bhardwaj}, M., {et~al.} 2020, arXiv
  e-prints, arXiv:2001.03595.
\newblock \doarXiv{2001.03595}

\bibitem[{{Gajjar} {et~al.}(2018){Gajjar}, {Siemion}, {Price}, {Law},
  {Michilli}, {Hessels}, {Chatterjee}, {Archibald}, {Bower}, {Brinkman},
  {Burke-Spolaor}, {Cordes}, {Croft}, {Enriquez}, {Foster}, {Gizani},
  {Hellbourg}, {Isaacson}, {Kaspi}, {Lazio}, {Lebofsky}, {Lynch}, {MacMahon},
  {McLaughlin}, {Ransom}, {Scholz}, {Seymour}, {Spitler}, {Tendulkar},
  {Werthimer}, \& {Zhang}}]{2018ApJ...863....2G}
{Gajjar}, V., {Siemion}, A.~P.~V., {Price}, D.~C., {et~al.} 2018, \apj, 863, 2,
  \dodoi{10.3847/1538-4357/aad005}

\bibitem[{{Gavriil} {et~al.}(2004){Gavriil}, {Kaspi}, \&
  {Woods}}]{2004ApJ...607..959G}
{Gavriil}, F.~P., {Kaspi}, V.~M., \& {Woods}, P.~M. 2004, \apj, 607, 959,
  \dodoi{10.1086/383564}

\bibitem[{{Goldreich} \& {Julian}(1969)}]{1969ApJ...157..869G}
{Goldreich}, P., \& {Julian}, W.~H. 1969, \apj, 157, 869,
  \dodoi{10.1086/150119}

\bibitem[{{Gourdji} {et~al.}(2019){Gourdji}, {Michilli}, {Spitler}, {Hessels},
  {Seymour}, {Cordes}, \& {Chatterjee}}]{2019ApJ...877L..19G}
{Gourdji}, K., {Michilli}, D., {Spitler}, L.~G., {et~al.} 2019, \apjl, 877,
  L19, \dodoi{10.3847/2041-8213/ab1f8a}

\bibitem[{{G{\"o}{\v g}{\"u}{\c s} } {et~al.}(1999){G{\"o}{\v g}{\"u}{\c s} },
  {Woods}, {Kouveliotou}, {van Paradijs}, {Briggs}, {Duncan}, \&
  {Thompson}}]{1999ApJ...526L..93G}
{G{\"o}{\v g}{\"u}{\c s} }, E., {Woods}, P.~M., {Kouveliotou}, C., {et~al.}
  1999, \apjl, 526, L93, \dodoi{10.1086/312380}

\bibitem[{{G{\"o}{\v g}{\"u}{\c s}} {et~al.}(2000){G{\"o}{\v g}{\"u}{\c s}},
  {Woods}, {Kouveliotou}, {van Paradijs}, {Briggs}, {Duncan}, \&
  {Thompson}}]{2000ApJ...532L.121G}
{G{\"o}{\v g}{\"u}{\c s}}, E., {Woods}, P.~M., {Kouveliotou}, C., {et~al.}
  2000, \apjl, 532, L121, \dodoi{10.1086/312583}

\bibitem[{{Harding} \& {Lai}(2006)}]{2006RPPh...69.2631H}
{Harding}, A.~K., \& {Lai}, D. 2006, Reports on Progress in Physics, 69, 2631,
  \dodoi{10.1088/0034-4885/69/9/R03}

\bibitem[{{Harding} \& {Muslimov}(1998)}]{1998ApJ...508..328H}
{Harding}, A.~K., \& {Muslimov}, A.~G. 1998, \apj, 508, 328,
  \dodoi{10.1086/306394}

\bibitem[{{Hassan} \& {Daniel}(2019)}]{2019ICRC...36..692H}
{Hassan}, T., \& {Daniel}, M. 2019, in International Cosmic Ray Conference,
  Vol.~36, 36th International Cosmic Ray Conference (ICRC2019), 692.
\newblock \doarXiv{1908.03393}

\bibitem[{{Hessels} {et~al.}(2019){Hessels}, {Spitler}, {Seymour}, {Cordes},
  {Michilli}, {Lynch}, {Gourdji}, {Archibald}, {Bassa}, {Bower}, {Chatterjee},
  {Connor}, {Crawford}, {Deneva}, {Gajjar}, {Kaspi}, {Keimpema}, {Law},
  {Marcote}, {McLaughlin}, {Paragi}, {Petroff}, {Ransom}, {Scholz}, {Stappers},
  \& {Tendulkar}}]{2019ApJ...876L..23H}
{Hessels}, J.~W.~T., {Spitler}, L.~G., {Seymour}, A.~D., {et~al.} 2019, The
  Astrophysical Journal, 876, L23, \dodoi{10.3847/2041-8213/ab13ae}

\bibitem[{{Hoang} {et~al.}(2019){Hoang}, {Will}, {Inoue}, {Barrio}, {Cortina},
  {Lopez}, {Marcote}, \& {Tejedor}}]{2019ICRC...36..697H}
{Hoang}, J., {Will}, M., {Inoue}, S., {et~al.} 2019, in International Cosmic
  Ray Conference, Vol.~36, 36th International Cosmic Ray Conference (ICRC2019),
  697.
\newblock \doarXiv{1908.07506}

\bibitem[{{Hoffman} \& {Heyl}(2012)}]{2012MNRAS.426.2404H}
{Hoffman}, K., \& {Heyl}, J. 2012, \mnras, 426, 2404,
  \dodoi{10.1111/j.1365-2966.2012.21921.x}

\bibitem[{{Horowitz} \& {Kadau}(2009)}]{2009PhRvL.102s1102H}
{Horowitz}, C.~J., \& {Kadau}, K. 2009, \prl, 102, 191102,
  \dodoi{10.1103/PhysRevLett.102.191102}

\bibitem[{{Hu} {et~al.}(2019){Hu}, {Baring}, {Wadiasingh}, \&
  {Harding}}]{2019MNRAS.tmp..982H}
{Hu}, K., {Baring}, M.~G., {Wadiasingh}, Z., \& {Harding}, A.~K. 2019, \mnras,
  \dodoi{10.1093/mnras/stz995}

\bibitem[{{Huppenkothen} {et~al.}(2014{\natexlab{a}}){Huppenkothen}, {Heil},
  {Watts}, \& {G{\"o}{\u g}{\"u}{\c s}}}]{2014ApJ...795..114H}
{Huppenkothen}, D., {Heil}, L.~M., {Watts}, A.~L., \& {G{\"o}{\u g}{\"u}{\c
  s}}, E. 2014{\natexlab{a}}, \apj, 795, 114,
  \dodoi{10.1088/0004-637X/795/2/114}

\bibitem[{{Huppenkothen} {et~al.}(2014{\natexlab{b}}){Huppenkothen}, {Watts},
  \& {Levin}}]{2014ApJ...793..129H}
{Huppenkothen}, D., {Watts}, A.~L., \& {Levin}, Y. 2014{\natexlab{b}}, \apj,
  793, 129, \dodoi{10.1088/0004-637X/793/2/129}

\bibitem[{{Huppenkothen} {et~al.}(2014{\natexlab{c}}){Huppenkothen},
  {D'Angelo}, {Watts}, {Heil}, {van der Klis}, {van der Horst}, {Kouveliotou},
  {Baring}, {G{\"o}{\u g}{\"u}{\c s}}, {Granot}, {Kaneko}, {Lin}, {von
  Kienlin}, \& {Younes}}]{2014ApJ...787..128H}
{Huppenkothen}, D., {D'Angelo}, C., {Watts}, A.~L., {et~al.}
  2014{\natexlab{c}}, \apj, 787, 128, \dodoi{10.1088/0004-637X/787/2/128}

\bibitem[{{Hurley} {et~al.}(1999){Hurley}, {Cline}, {Mazets}, {Barthelmy},
  {Butterworth}, {Marshall}, {Palmer}, {Aptekar}, {Golenetskii}, {Il'Inskii},
  {Frederiks}, {McTiernan}, {Gold}, \& {Trombka}}]{1999Natur.397...41H}
{Hurley}, K., {Cline}, T., {Mazets}, E., {et~al.} 1999, \nat, 397, 41,
  \dodoi{10.1038/16199}

\bibitem[{{Israel} {et~al.}(2008){Israel}, {Romano}, {Mangano}, {Dall'Osso},
  {Chincarini}, {Stella}, {Campana}, {Belloni}, {Tagliaferri}, {Blustin},
  {Sakamoto}, {Hurley}, {Zane}, {Moretti}, {Palmer}, {Guidorzi}, {Burrows},
  {Gehrels}, \& {Krimm}}]{2008ApJ...685.1114I}
{Israel}, G.~L., {Romano}, P., {Mangano}, V., {et~al.} 2008, \apj, 685, 1114,
  \dodoi{10.1086/590486}

\bibitem[{{James} {et~al.}(2019){James}, {Oslowski}, {Flynn}, {Kumar},
  {Bannister}, {Bhandari}, {Farah}, {Kerr}, {Lorimer}, {Macquart}, {Ng},
  {Phillips}, {Price}, {Qiu}, {Shannon}, \& {Spiewak}}]{2019arXiv191207847J}
{James}, C.~W., {Oslowski}, S., {Flynn}, C., {et~al.} 2019, arXiv e-prints,
  arXiv:1912.07847.
\newblock \doarXiv{1912.07847}

\bibitem[{{Katz}(2016)}]{2016ApJ...826..226K}
{Katz}, J.~I. 2016, \apj, 826, 226, \dodoi{10.3847/0004-637X/826/2/226}

\bibitem[{{Katz}(2018)}]{2018PrPNP.103....1K}
---. 2018, Progress in Particle and Nuclear Physics, 103, 1,
  \dodoi{10.1016/j.ppnp.2018.07.001}

\bibitem[{{Kumar} {et~al.}(2010){Kumar}, {Ibrahim}, \&
  {Safi-Harb}}]{2010ApJ...716...97K}
{Kumar}, H.~S., {Ibrahim}, A.~I., \& {Safi-Harb}, S. 2010, \apj, 716, 97,
  \dodoi{10.1088/0004-637X/716/1/97}

\bibitem[{{Kumar} {et~al.}(2017){Kumar}, {Lu}, \&
  {Bhattacharya}}]{2017MNRAS.468.2726K}
{Kumar}, P., {Lu}, W., \& {Bhattacharya}, M. 2017, \mnras, 468, 2726,
  \dodoi{10.1093/mnras/stx665}

\bibitem[{{Kumar} {et~al.}(2019){Kumar}, {Shannon}, {Os{\l}owski}, {Qiu},
  {Bhandari}, {Farah}, {Flynn}, {Kerr}, {Lorimer}, {Macquart}, {Ng},
  {Phillips}, {Price}, \& {Spiewak}}]{2019arXiv190810026K}
{Kumar}, P., {Shannon}, R.~M., {Os{\l}owski}, S., {et~al.} 2019, arXiv
  e-prints, arXiv:1908.10026.
\newblock \doarXiv{1908.10026}

\bibitem[{{Kundu} \& {Ferrario}(2019)}]{2019arXiv191208287K}
{Kundu}, E., \& {Ferrario}, L. 2019, arXiv e-prints, arXiv:1912.08287.
\newblock \doarXiv{1912.08287}

\bibitem[{{Lacki}(2011)}]{2011MNRAS.416.3075L}
{Lacki}, B.~C. 2011, \mnras, 416, 3075,
  \dodoi{10.1111/j.1365-2966.2011.19255.x}

\bibitem[{{Lai}(2001)}]{2001RvMP...73..629L}
{Lai}, D. 2001, Reviews of Modern Physics, 73, 629,
  \dodoi{10.1103/RevModPhys.73.629}

\bibitem[{{Lai} \& {Ho}(2002)}]{2002ApJ...566..373L}
{Lai}, D., \& {Ho}, W.~C.~G. 2002, \apj, 566, 373, \dodoi{10.1086/338074}

\bibitem[{{Lai} \& {Salpeter}(1997)}]{1997ApJ...491..270L}
{Lai}, D., \& {Salpeter}, E.~E. 1997, \apj, 491, 270, \dodoi{10.1086/304937}

\bibitem[{{Lander} {et~al.}(2015){Lander}, {Andersson}, {Antonopoulou}, \&
  {Watts}}]{2015MNRAS.449.2047L}
{Lander}, S.~K., {Andersson}, N., {Antonopoulou}, D., \& {Watts}, A.~L. 2015,
  \mnras, 449, 2047, \dodoi{10.1093/mnras/stv432}

\bibitem[{{Levin}(2006)}]{2006MNRAS.368L..35L}
{Levin}, Y. 2006, \mnras, 368, L35, \dodoi{10.1111/j.1745-3933.2006.00155.x}

\bibitem[{{Lin} \& {Sang}(2019)}]{2019MNRAS.tmp.2746L}
{Lin}, H.-N., \& {Sang}, Y. 2019, \mnras, 2746, \dodoi{10.1093/mnras/stz3149}

\bibitem[{{Lin} {et~al.}(2011){Lin}, {Kouveliotou}, {Baring}, {van der Horst},
  {Guiriec}, {Woods}, {G{\"o}{\v{g}}{\"u}{\textcommabelow s}}, {Kaneko},
  {Scargle}, {Granot}, {Preece}, {von Kienlin}, {Chaplin}, {Watts}, {Wijers},
  {Zhang}, {Bhat}, {Finger}, {Gehrels}, {Harding}, {Kaper}, {Kaspi}, {Mcenery},
  {Meegan}, {Paciesas}, {Pe'er}, {Ramirez-Ruiz}, {van der Klis}, {Wachter}, \&
  {Wilson-Hodge}}]{2011ApJ...739...87L}
{Lin}, L., {Kouveliotou}, C., {Baring}, M.~G., {et~al.} 2011, \apj, 739, 87,
  \dodoi{10.1088/0004-637X/739/2/87}

\bibitem[{{Lin} {et~al.}(2012){Lin}, {G{\"o}{\v g}{\"u}{\c s}}, {Baring},
  {Granot}, {Kouveliotou}, {Kaneko}, {van der Horst}, {Gruber}, {von Kienlin},
  {Younes}, {Watts}, \& {Gehrels}}]{2012ApJ...756...54L}
{Lin}, L., {G{\"o}{\v g}{\"u}{\c s}}, E., {Baring}, M.~G., {et~al.} 2012, \apj,
  756, 54, \dodoi{10.1088/0004-637X/756/1/54}

\bibitem[{{Linscott} \& {Erkes}(1980)}]{1980ApJ...236L.109L}
{Linscott}, I.~R., \& {Erkes}, J.~W. 1980, \apjl, 236, L109,
  \dodoi{10.1086/183209}

\bibitem[{{Luo} {et~al.}(2018){Luo}, {Lee}, {Lorimer}, \&
  {Zhang}}]{2018MNRAS.481.2320L}
{Luo}, R., {Lee}, K., {Lorimer}, D.~R., \& {Zhang}, B. 2018, \mnras, 481, 2320,
  \dodoi{10.1093/mnras/sty2364}

\bibitem[{{Lyubarskii} \& {Petrova}(1998)}]{1998Ap&SS.262..379L}
{Lyubarskii}, Y.~E., \& {Petrova}, S.~A. 1998, \apss, 262, 379,
  \dodoi{10.1023/A:1001872805645}

\bibitem[{{Lyubarsky}(2014)}]{2014MNRAS.442L...9L}
{Lyubarsky}, Y. 2014, \mnras, 442, L9, \dodoi{10.1093/mnrasl/slu046}

\bibitem[{{Lyubarsky}(2020)}]{2020arXiv200102007L}
---. 2020, arXiv e-prints, arXiv:2001.02007.
\newblock \doarXiv{2001.02007}

\bibitem[{{Lyutikov}(2015)}]{2015MNRAS.447.1407L}
{Lyutikov}, M. 2015, \mnras, 447, 1407, \dodoi{10.1093/mnras/stu2413}

\bibitem[{{Lyutikov}(2017)}]{2017ApJ...838L..13L}
---. 2017, \apjl, 838, L13, \dodoi{10.3847/2041-8213/aa62fa}

\bibitem[{{Lyutikov}(2019{\natexlab{a}})}]{2019arXiv190103260L}
---. 2019{\natexlab{a}}, arXiv e-prints.
\newblock \doarXiv{1901.03260}

\bibitem[{{Lyutikov}(2019{\natexlab{b}})}]{2019arXiv190807313L}
---. 2019{\natexlab{b}}, arXiv e-prints, arXiv:1908.07313.
\newblock \doarXiv{1908.07313}

\bibitem[{{Madau} \& {Dickinson}(2014)}]{2014ARA&A..52..415M}
{Madau}, P., \& {Dickinson}, M. 2014, \araa, 52, 415,
  \dodoi{10.1146/annurev-astro-081811-125615}

\bibitem[{{MAGIC Collaboration} {et~al.}(2018){MAGIC Collaboration}, {Acciari},
  {Ansoldi}, {Antonelli}, {Arbet Engels}, {Arcaro}, {Baack}, {Babi{\'c}}, {},
  {Banerjee}, {Bangale}, {Barres de Almeida}, {Barrio}, {Becerra Gonz{\'a}lez},
  {Bednarek}, {Bernardini}, {Berti}, {Besenrieder}, {Bhattacharyya},
  {Bigongiari}, {Biland}, {Blanch}, {Bonnoli}, {Carosi}, {Ceribella},
  {Chatterjee}, {Colak}, {Colin}, {Colombo}, {Contreras}, {Cortina}, {Covino},
  {Cumani}, {D'Elia}, {da Vela}, {Dazzi}, {de Angelis}, {de Lotto}, {Delfino},
  {Delgado}, {di Pierro}, {Dom{\'\i}nguez}, {Dominis Prester}, {Dorner},
  {Doro}, {Einecke}, {Elsaesser}, {Fallah Ramazani}, {Fattorini},
  {Fern{\'a}ndez-Barral}, {Ferrara}, {Fidalgo}, {Foffano}, {Fonseca}, {Font},
  {Fruck}, {Gallozzi}, {Garc{\'\i}a L{\'o}pez}, {Garczarczyk}, {Gaug},
  {Giammaria}, {Godinovi{\'c}}, {}, {Guberman}, {Hadasch}, {Hahn}, {Hassan},
  {Herrera}, {Hoang}, {Hrupec}, {Inoue}, {Ishio}, {Iwamura}, {Kubo}, {Kushida},
  {Kuve{\v{z}}di{\'c}}, {}, {Lamastra}, {Lelas}, {Leone}, {Lindfors},
  {Lombardi}, {Longo}, {L{\'o}pez}, {L{\'o}pez-Oramas}, {Maggio}, {Majumdar},
  {Makariev}, {Maneva}, {Manganaro}, {Mannheim}, {Maraschi}, {Mariotti},
  {Mart{\'\i}nez}, {Masuda}, {Mazin}, {Minev}, {Miranda}, {Mirzoyan}, {Molina},
  {Moralejo}, {Moreno}, {Moretti}, {Neustroev}, {Niedzwiecki}, {Nievas
  Rosillo}, {Nigro}, {Nilsson}, {Ninci}, {Nishijima}, {Noda}, {Nogu{\'e}s},
  {Paiano}, {Palacio}, {Paneque}, {Paoletti}, {Paredes}, {Pedaletti},
  {Pe{\~n}il}, {Peresano}, {Persic}, {Prada Moroni}, {Prand ini}, {Puljak},
  {Garcia}, {Rhode}, {Rib{\'o}}, {Rico}, {Righi}, {Rugliancich}, {Saha},
  {Saito}, {Satalecka}, {Schweizer}, {Sitarek}, {{\v{S}}nidari{\'c}}, {},
  {Sobczynska}, {Somero}, {Stamerra}, {Strzys}, {Suri{\'c}}, {}, {Tavecchio},
  {Temnikov}, {Terzi{\'c}}, {}, {Teshima}, {Torres-Alb{\`a}}, {Tsujimoto},
  {Vanzo}, {Vazquez Acosta}, {Vovk}, {Ward}, {Will}, {Zari{\'c}}, {Marcote},
  {Spitler}, {Hessels}, {Kashiyama}, {Murase}, {Bosch-Ramon}, {Michilli}, \&
  {Seymour}}]{2018MNRAS.481.2479M}
{MAGIC Collaboration}, {Acciari}, V.~A., {Ansoldi}, S., {et~al.} 2018, \mnras,
  481, 2479, \dodoi{10.1093/mnras/sty2422}

\bibitem[{{Mahony} {et~al.}(2018){Mahony}, {Ekers}, {Macquart}, {Sadler},
  {Bannister}, {Bhandari}, {Flynn}, {Koribalski}, {Prochaska}, {Ryder},
  {Shannon}, {Tejos}, {Whiting}, \& {Wong}}]{2018ApJ...867L..10M}
{Mahony}, E.~K., {Ekers}, R.~D., {Macquart}, J.-P., {et~al.} 2018, \apjl, 867,
  L10, \dodoi{10.3847/2041-8213/aae7cb}

\bibitem[{{Manchester} {et~al.}(2005){Manchester}, {Hobbs}, {Teoh}, \&
  {Hobbs}}]{2005AJ....129.1993M}
{Manchester}, R.~N., {Hobbs}, G.~B., {Teoh}, A., \& {Hobbs}, M. 2005, \aj, 129,
  1993, \dodoi{10.1086/428488}

\bibitem[{Marcote {et~al.}(2020)Marcote, Nimmo, Hessels, Tendulkar, Bassa,
  Paragi, Keimpema, Bhardwaj, Karuppusamy, Kaspi, Law, Michilli, Aggarwal,
  Andersen, Archibald, Bandura, Bower, Boyle, Brar, Burke-Spolaor, Butler,
  Cassanelli, Chawla, Demorest, Dobbs, Fonseca, Giri, Good, Gourdji, Josephy,
  Kirichenko, Kirsten, Landecker, Lang, Lazio, Li, Lin, Linford, Masui,
  Mena-Parra, Naidu, Ng, Patel, Pen, Pleunis, Rafiei-Ravandi, Rahman, Renard,
  Scholz, Siegel, Smith, Stairs, Vanderlinde, \& Zwaniga}]{Marcote2020}
Marcote, B., Nimmo, K., Hessels, J. W.~T., {et~al.} 2020, Nature,
  \dodoi{10.1038/s41586-019-1866-z}

\bibitem[{{Medin} \& {Lai}(2006)}]{2006PhRvA..74f2508M}
{Medin}, Z., \& {Lai}, D. 2006, \pra, 74, 062508,
  \dodoi{10.1103/PhysRevA.74.062508}

\bibitem[{{Medin} \& {Lai}(2007)}]{2007MNRAS.382.1833M}
---. 2007, \mnras, 382, 1833, \dodoi{10.1111/j.1365-2966.2007.12492.x}

\bibitem[{{Melrose}(1979)}]{1979AuJPh..32...61M}
{Melrose}, D.~B. 1979, Australian Journal of Physics, 32, 61,
  \dodoi{10.1071/PH790061}

\bibitem[{{Melrose}(2017)}]{2017RvMPP...1....5M}
---. 2017, Reviews of Modern Plasma Physics, 1, 5,
  \dodoi{10.1007/s41614-017-0007-0}

\bibitem[{{Metzger} {et~al.}(2019){Metzger}, {Margalit}, \&
  {Sironi}}]{2019MNRAS.485.4091M}
{Metzger}, B.~D., {Margalit}, B., \& {Sironi}, L. 2019, \mnras, 485, 4091,
  \dodoi{10.1093/mnras/stz700}

\bibitem[{{Michilli} {et~al.}(2018){Michilli}, {Seymour}, {Hessels}, {Spitler},
  {Gajjar}, {Archibald}, {Bower}, {Chatterjee}, {Cordes}, {Gourdji}, {Heald},
  {Kaspi}, {Law}, {Sobey}, {Adams}, {Bassa}, {Bogdanov}, {Brinkman},
  {Demorest}, {Fernandez}, {Hellbourg}, {Lazio}, {Lynch}, {Maddox}, {Marcote},
  {McLaughlin}, {Paragi}, {Ransom}, {Scholz}, {Siemion}, {Tendulkar}, {van
  Rooy}, {Wharton}, \& {Whitlow}}]{2018Natur.553..182M}
{Michilli}, D., {Seymour}, A., {Hessels}, J.~W.~T., {et~al.} 2018, \nat, 553,
  182, \dodoi{10.1038/nature25149}

\bibitem[{{Miller} {et~al.}(2019){Miller}, {Chirenti}, \&
  {Strohmayer}}]{2019ApJ...871...95M}
{Miller}, M.~C., {Chirenti}, C., \& {Strohmayer}, T.~E. 2019, \apj, 871, 95,
  \dodoi{10.3847/1538-4357/aaf5ce}

\bibitem[{{Morello} {et~al.}(2019){Morello}, {Keane}, {Enoto}, {Guillot}, {Ho},
  {Jameson}, {Kramer}, {Stappers}, {Bailes}, {Barr}, {Bhand ari}, {Caleb},
  {Flynn}, {Jankowski}, {Johnston}, {van Straten}, {Arzoumanian}, {Bogdanov},
  {Gendreau}, {Malacaria}, {Ray}, \& {Remillard}}]{2019arXiv191004124M}
{Morello}, V., {Keane}, E.~F., {Enoto}, T., {et~al.} 2019, arXiv e-prints,
  arXiv:1910.04124.
\newblock \doarXiv{1910.04124}

\bibitem[{{Murase} {et~al.}(2016){Murase}, {Kashiyama}, \&
  {M{\'e}sz{\'a}ros}}]{2016MNRAS.461.1498M}
{Murase}, K., {Kashiyama}, K., \& {M{\'e}sz{\'a}ros}, P. 2016, \mnras, 461,
  1498, \dodoi{10.1093/mnras/stw1328}

\bibitem[{{Olausen} \& {Kaspi}(2014)}]{2014ApJS..212....6O}
{Olausen}, S.~A., \& {Kaspi}, V.~M. 2014, \apjs, 212, 6,
  \dodoi{10.1088/0067-0049/212/1/6}

\bibitem[{{Oostrum} {et~al.}(2019){Oostrum}, {Maan}, {van Leeuwen}, {Connor},
  {Petroff}, {Attema}, {Bast}, {Gardenier}, {Hargreaves}, {Kooistra},
  {Sclocco}, {Smits}, {Straal}, {ter Veen}, {Vohl}, {Adams}, {Adebahr}, {de
  Blok}, {van den Brink}, {van Cappellen}, {Coolen}, {Damstra}, {van Diepen},
  {Frank}, {Hess}, {van der Hulst}, {Hut}, {Ivashina}, {Loose}, {Lucero},
  {Mika}, {Morganti}, {Moss}, {Mulder}, {Norden}, {Oosterloo}, {Orr{\'u}}, {de
  Reijer}, {Ruiter}, {Vermaas}, {Wijnholds}, \& {Ziemke}}]{2019arXiv191212217O}
{Oostrum}, L.~C., {Maan}, Y., {van Leeuwen}, J., {et~al.} 2019, arXiv e-prints,
  arXiv:1912.12217.
\newblock \doarXiv{1912.12217}

\bibitem[{{Palaniswamy} {et~al.}(2018){Palaniswamy}, {Li}, \&
  {Zhang}}]{2018ApJ...854L..12P}
{Palaniswamy}, D., {Li}, Y., \& {Zhang}, B. 2018, \apjl, 854, L12,
  \dodoi{10.3847/2041-8213/aaaa63}

\bibitem[{{Perna} \& {Pons}(2011)}]{2011ApJ...727L..51P}
{Perna}, R., \& {Pons}, J.~A. 2011, \apj, 727, L51,
  \dodoi{10.1088/2041-8205/727/2/L51}

\bibitem[{{Petroff} {et~al.}(2019){Petroff}, {Hessels}, \&
  {Lorimer}}]{2019A&ARv..27....4P}
{Petroff}, E., {Hessels}, J.~W.~T., \& {Lorimer}, D.~R. 2019, \aapr, 27, 4,
  \dodoi{10.1007/s00159-019-0116-6}

\bibitem[{{Petrova} \& {Lyubarskii}(2000)}]{2000A&A...355.1168P}
{Petrova}, S.~A., \& {Lyubarskii}, Y.~E. 2000, \aap, 355, 1168

\bibitem[{{Philippov} {et~al.}(2020){Philippov}, {Timokhin}, \&
  {Spitkovsky}}]{2020arXiv200102236P}
{Philippov}, A., {Timokhin}, A., \& {Spitkovsky}, A. 2020, arXiv e-prints,
  arXiv:2001.02236.
\newblock \doarXiv{2001.02236}

\bibitem[{{Planck Collaboration} {et~al.}(2016){Planck Collaboration}, {Ade},
  {Aghanim}, {Arnaud}, {Ashdown}, {Aumont}, {Baccigalupi}, {Banday},
  {Barreiro}, {Bartlett}, {Bartolo}, {Battaner}, {Battye}, {Benabed},
  {Beno{\^\i}t}, {Benoit-L{\'e}vy}, {Bernard}, {Bersanelli}, {Bielewicz},
  {Bock}, {Bonaldi}, {Bonavera}, {Bond}, {Borrill}, {Bouchet}, {Boulanger},
  {Bucher}, {Burigana}, {Butler}, {Calabrese}, {Cardoso}, {Catalano},
  {Challinor}, {Chamballu}, {Chary}, {Chiang}, {Chluba}, {Christensen},
  {Church}, {Clements}, {Colombi}, {Colombo}, {Combet}, {Coulais}, {Crill},
  {Curto}, {Cuttaia}, {Danese}, {Davies}, {Davis}, {de Bernardis}, {de Rosa},
  {de Zotti}, {Delabrouille}, {D{\'e}sert}, {Di Valentino}, {Dickinson},
  {Diego}, {Dolag}, {Dole}, {Donzelli}, {Dor{\'e}}, {Douspis}, {Ducout},
  {Dunkley}, {Dupac}, {Efstathiou}, {Elsner}, {En{\ss}lin}, {Eriksen},
  {Farhang}, {Fergusson}, {Finelli}, {Forni}, {Frailis}, {Fraisse},
  {Franceschi}, {Frejsel}, {Galeotta}, {Galli}, {Ganga}, {Gauthier}, {Gerbino},
  {Ghosh}, {Giard}, {Giraud-H{\'e}raud}, {Giusarma}, {Gjerl{\o}w},
  {Gonz{\'a}lez-Nuevo}, {G{\'o}rski}, {Gratton}, {Gregorio}, {Gruppuso},
  {Gudmundsson}, {Hamann}, {Hansen}, {Hanson}, {Harrison}, {Helou},
  {Henrot-Versill{\'e}}, {Hern{\'a}ndez-Monteagudo}, {Herranz}, {Hildebrand t},
  {Hivon}, {Hobson}, {Holmes}, {Hornstrup}, {Hovest}, {Huang}, {Huffenberger},
  {Hurier}, {Jaffe}, {Jaffe}, {Jones}, {Juvela}, {Keih{\"a}nen}, {Keskitalo},
  {Kisner}, {Kneissl}, {Knoche}, {Knox}, {Kunz}, {Kurki-Suonio}, {Lagache},
  {L{\"a}hteenm{\"a}ki}, {Lamarre}, {Lasenby}, {Lattanzi}, {Lawrence}, {Leahy},
  {Leonardi}, {Lesgourgues}, {Levrier}, {Lewis}, {Liguori}, {Lilje},
  {Linden-V{\o}rnle}, {L{\'o}pez-Caniego}, {Lubin}, {Mac{\'\i}as-P{\'e}rez},
  {Maggio}, {Maino}, {Mandolesi}, {Mangilli}, {Marchini}, {Maris}, {Martin},
  {Martinelli}, {Mart{\'\i}nez-Gonz{\'a}lez}, {Masi}, {Matarrese}, {McGehee},
  {Meinhold}, {Melchiorri}, {Melin}, {Mendes}, {Mennella}, {Migliaccio},
  {Millea}, {Mitra}, {Miville-Desch{\^e}nes}, {Moneti}, {Montier}, {Morgante},
  {Mortlock}, {Moss}, {Munshi}, {Murphy}, {Naselsky}, {Nati}, {Natoli},
  {Netterfield}, {N{\o}rgaard-Nielsen}, {Noviello}, {Novikov}, {Novikov},
  {Oxborrow}, {Paci}, {Pagano}, {Pajot}, {Paladini}, {Paoletti}, {Partridge},
  {Pasian}, {Patanchon}, {Pearson}, {Perdereau}, {Perotto}, {Perrotta},
  {Pettorino}, {Piacentini}, {Piat}, {Pierpaoli}, {Pietrobon}, {Plaszczynski},
  {Pointecouteau}, {Polenta}, {Popa}, {Pratt}, {Pr{\'e}zeau}, {Prunet},
  {Puget}, {Rachen}, {Reach}, {Rebolo}, {Reinecke}, {Remazeilles}, {Renault},
  {Renzi}, {Ristorcelli}, {Rocha}, {Rosset}, {Rossetti}, {Roudier},
  {Rouill{\'e} d'Orfeuil}, {Rowan-Robinson}, {Rubi{\~n}o-Mart{\'\i}n},
  {Rusholme}, {Said}, {Salvatelli}, {Salvati}, {Sandri}, {Santos},
  {Savelainen}, {Savini}, {Scott}, {Seiffert}, {Serra}, {Shellard}, {Spencer},
  {Spinelli}, {Stolyarov}, {Stompor}, {Sudiwala}, {Sunyaev}, {Sutton},
  {Suur-Uski}, {Sygnet}, {Tauber}, {Terenzi}, {Toffolatti}, {Tomasi},
  {Tristram}, {Trombetti}, {Tucci}, {Tuovinen}, {T{\"u}rler}, {Umana},
  {Valenziano}, {Valiviita}, {Van Tent}, {Vielva}, {Villa}, {Wade}, {Wandelt},
  {Wehus}, {White}, {White}, {Wilkinson}, {Yvon}, {Zacchei}, \&
  {Zonca}}]{2016A&A...594A..13P}
{Planck Collaboration}, {Ade}, P.~A.~R., {Aghanim}, N., {et~al.} 2016, \aap,
  594, A13, \dodoi{10.1051/0004-6361/201525830}

\bibitem[{{Platts} {et~al.}(2018){Platts}, {Weltman}, {Walters}, {Tendulkar},
  {Gordin}, \& {Kandhai}}]{2018arXiv181005836P}
{Platts}, E., {Weltman}, A., {Walters}, A., {et~al.} 2018, arXiv e-prints,
  arXiv:1810.05836.
\newblock \doarXiv{1810.05836}

\bibitem[{{Popov} \& {Postnov}(2010)}]{2010vaoa.conf..129P}
{Popov}, S.~B., \& {Postnov}, K.~A. 2010, in Evolution of Cosmic Objects
  through their Physical Activity, ed. H.~A. {Harutyunian}, A.~M. {Mickaelian},
  \& Y.~{Terzian}, 129--132.
\newblock \doarXiv{0710.2006}

\bibitem[{{Popov} \& {Postnov}(2013)}]{2013arXiv1307.4924P}
{Popov}, S.~B., \& {Postnov}, K.~A. 2013, arXiv e-prints, arXiv:1307.4924.
\newblock \doarXiv{1307.4924}

\bibitem[{{Prieskorn} \& {Kaaret}(2012)}]{2012ApJ...755....1P}
{Prieskorn}, Z., \& {Kaaret}, P. 2012, \apj, 755, 1,
  \dodoi{10.1088/0004-637X/755/1/1}

\bibitem[{Prochaska {et~al.}(2019)Prochaska, Macquart, McQuinn, Simha, Shannon,
  Day, Marnoch, Ryder, Deller, Bannister, Bhandari, Bordoloi, Bunton, Cho,
  Flynn, Mahony, Phillips, Qiu, \& Tejos}]{Prochaska231}
Prochaska, J.~X., Macquart, J.-P., McQuinn, M., {et~al.} 2019, Science, 366,
  231, \dodoi{10.1126/science.aay0073}

\bibitem[{{Ramaty} {et~al.}(1980){Ramaty}, {Bonazzola}, {Cline}, {Kazanas},
  {Meszaros}, \& {Lingenfelter}}]{1980Natur.287..122R}
{Ramaty}, R., {Bonazzola}, S., {Cline}, T.~L., {et~al.} 1980, \nat, 287, 122,
  \dodoi{10.1038/287122a0}

\bibitem[{{Ravi}(2019)}]{2019NatAs...3..928R}
{Ravi}, V. 2019, Nature Astronomy, 3, 928, \dodoi{10.1038/s41550-019-0831-y}

\bibitem[{{Ravi} {et~al.}(2019){Ravi}, {Catha}, {D'Addario}, {Djorgovski},
  {Hallinan}, {Hobbs}, {Kocz}, {Kulkarni}, {Shi}, {Vedantham}, {Weinreb}, \&
  {Woody}}]{2019Natur.572..352R}
{Ravi}, V., {Catha}, M., {D'Addario}, L., {et~al.} 2019, \nat, 572, 352,
  \dodoi{10.1038/s41586-019-1389-7}

\bibitem[{{Rea} {et~al.}(2016){Rea}, {Borghese}, {Esposito}, {Coti Zelati},
  {Bachetti}, {Israel}, \& {De Luca}}]{2016ApJ...828L..13R}
{Rea}, N., {Borghese}, A., {Esposito}, P., {et~al.} 2016, \apjl, 828, L13,
  \dodoi{10.3847/2041-8205/828/1/L13}

\bibitem[{{Ruderman} \& {Sutherland}(1975)}]{1975ApJ...196...51R}
{Ruderman}, M.~A., \& {Sutherland}, P.~G. 1975, \apj, 196, 51,
  \dodoi{10.1086/153393}

\bibitem[{{Savchenko} {et~al.}(2010){Savchenko}, {Neronov}, {Beckmann},
  {Produit}, \& {Walter}}]{2010A&A...510A..77S}
{Savchenko}, V., {Neronov}, A., {Beckmann}, V., {Produit}, N., \& {Walter}, R.
  2010, \aap, 510, A77, \dodoi{10.1051/0004-6361/200911988}

\bibitem[{{Scholz} \& {Kaspi}(2011)}]{2011ApJ...739...94S}
{Scholz}, P., \& {Kaspi}, V.~M. 2011, \apj, 739, 94,
  \dodoi{10.1088/0004-637X/739/2/94}

\bibitem[{{Shannon} {et~al.}(2018){Shannon}, {Macquart}, {Bannister}, {Ekers},
  {James}, {Os{\l}owski}, {Qiu}, {Sammons}, {Hotan}, {Voronkov}, {Beresford},
  {Brothers}, {Brown}, {Bunton}, {Chippendale}, {Haskins}, {Leach},
  {Marquarding}, {McConnell}, {Pilawa}, {Sadler}, {Troup}, {Tuthill},
  {Whiting}, {Allison}, {Anderson}, {Bell}, {Collier}, {G{\"u}rkan}, {Heald},
  \& {Riseley}}]{2018Natur.562..386S}
{Shannon}, R.~M., {Macquart}, J.~P., {Bannister}, K.~W., {et~al.} 2018, \nat,
  562, 386, \dodoi{10.1038/s41586-018-0588-y}

\bibitem[{{Spitler} {et~al.}(2016){Spitler}, {Scholz}, {Hessels}, {Bogdanov},
  {Brazier}, {Camilo}, {Chatterjee}, {Cordes}, {Crawford}, {Deneva}, {Ferdman},
  {Freire}, {Kaspi}, {Lazarus}, {Lynch}, {Madsen}, {McLaughlin}, {Patel},
  {Ransom}, {Seymour}, {Stairs}, {Stappers}, {van Leeuwen}, \&
  {Zhu}}]{2016Natur.531..202S}
{Spitler}, L.~G., {Scholz}, P., {Hessels}, J.~W.~T., {et~al.} 2016, \nat, 531,
  202, \dodoi{10.1038/nature17168}

\bibitem[{{Steiner} \& {Watts}(2009)}]{2009PhRvL.103r1101S}
{Steiner}, A.~W., \& {Watts}, A.~L. 2009, \prl, 103, 181101,
  \dodoi{10.1103/PhysRevLett.103.181101}

\bibitem[{{Story} \& {Baring}(2014)}]{2014ApJ...790...61S}
{Story}, S.~A., \& {Baring}, M.~G. 2014, \apj, 790, 61,
  \dodoi{10.1088/0004-637X/790/1/61}

\bibitem[{{Sturrock}(1971)}]{1971ApJ...164..529S}
{Sturrock}, P.~A. 1971, \apj, 164, 529, \dodoi{10.1086/150865}

\bibitem[{{Suvorov} \& {Kokkotas}(2019)}]{2019arXiv190710394S}
{Suvorov}, A.~G., \& {Kokkotas}, K.~D. 2019, arXiv e-prints, arXiv:1907.10394.
\newblock \doarXiv{1907.10394}

\bibitem[{{Tan} {et~al.}(2018){Tan}, {Bassa}, {Cooper}, {Dijkema}, {Esposito},
  {Hessels}, {Kondratiev}, {Kramer}, {Michilli}, {Sanidas}, {Shimwell},
  {Stappers}, {van Leeuwen}, {Cognard}, {Grie{\ss}meier}, {Karastergiou},
  {Keane}, {Sobey}, \& {Weltevrede}}]{2018ApJ...866...54T}
{Tan}, C.~M., {Bassa}, C.~G., {Cooper}, S., {et~al.} 2018, \apj, 866, 54,
  \dodoi{10.3847/1538-4357/aade88}

\bibitem[{{Tendulkar} {et~al.}(2017){Tendulkar}, {Bassa}, {Cordes}, {Bower},
  {Law}, {Chatterjee}, {Adams}, {Bogdanov}, {Burke-Spolaor}, {Butler},
  {Demorest}, {Hessels}, {Kaspi}, {Lazio}, {Maddox}, {Marcote}, {McLaughlin},
  {Paragi}, {Ransom}, {Scholz}, {Seymour}, {Spitler}, {van Langevelde}, \&
  {Wharton}}]{2017ApJ...834L...7T}
{Tendulkar}, S.~P., {Bassa}, C.~G., {Cordes}, J.~M., {et~al.} 2017, \apjl, 834,
  L7, \dodoi{10.3847/2041-8213/834/2/L7}

\bibitem[{{Thompson} \& {Duncan}(1995)}]{1995MNRAS.275..255T}
{Thompson}, C., \& {Duncan}, R.~C. 1995, \mnras, 275, 255,
  \dodoi{10.1093/mnras/275.2.255}

\bibitem[{{Thompson} \& {Duncan}(2001)}]{2001ApJ...561..980T}
---. 2001, \apj, 561, 980, \dodoi{10.1086/323256}

\bibitem[{{Tiengo} {et~al.}(2013){Tiengo}, {Esposito}, {Mereghetti}, {Turolla},
  {Nobili}, {Gastaldello}, {G{\"o}tz}, {Israel}, {Rea}, {Stella}, {Zane}, \&
  {Bignami}}]{2013Natur.500..312T}
{Tiengo}, A., {Esposito}, P., {Mereghetti}, S., {et~al.} 2013, \nat, 500, 312,
  \dodoi{10.1038/nature12386}

\bibitem[{{Timokhin}(2010)}]{2010MNRAS.408.2092T}
{Timokhin}, A.~N. 2010, \mnras, 408, 2092,
  \dodoi{10.1111/j.1365-2966.2010.17286.x}

\bibitem[{{Timokhin} \& {Arons}(2013)}]{2013MNRAS.429...20T}
{Timokhin}, A.~N., \& {Arons}, J. 2013, \mnras, 429, 20,
  \dodoi{10.1093/mnras/sts298}

\bibitem[{{Timokhin} \& {Harding}(2015)}]{2015ApJ...810..144T}
{Timokhin}, A.~N., \& {Harding}, A.~K. 2015, \apj, 810, 144,
  \dodoi{10.1088/0004-637X/810/2/144}

\bibitem[{{Timokhin} \& {Harding}(2019)}]{2019ApJ...871...12T}
---. 2019, \apj, 871, 12, \dodoi{10.3847/1538-4357/aaf050}

\bibitem[{{Usov}(1987)}]{1987ApJ...320..333U}
{Usov}, V.~V. 1987, \apj, 320, 333, \dodoi{10.1086/165546}

\bibitem[{{van der Horst} {et~al.}(2012){van der Horst}, {Kouveliotou},
  {Gorgone}, {Kaneko}, {Baring}, {Guiriec}, {G{\"o}{\v g}{\"u}{\c s}},
  {Granot}, {Watts}, {Lin}, {Bhat}, {Bissaldi}, {Chaplin}, {Finger}, {Gehrels},
  {Gibby}, {Giles}, {Goldstein}, {Gruber}, {Harding}, {Kaper}, {von Kienlin},
  {van der Klis}, {McBreen}, {Mcenery}, {Meegan}, {Paciesas}, {Pe'er},
  {Preece}, {Ramirez-Ruiz}, {Rau}, {Wachter}, {Wilson-Hodge}, {Woods}, \&
  {Wijers}}]{2012ApJ...749..122V}
{van der Horst}, A.~J., {Kouveliotou}, C., {Gorgone}, N.~M., {et~al.} 2012,
  \apj, 749, 122, \dodoi{10.1088/0004-637X/749/2/122}

\bibitem[{{van Putten} {et~al.}(2016){van Putten}, {Watts}, {Baring}, \&
  {Wijers}}]{2016MNRAS.461..877V}
{van Putten}, T., {Watts}, A.~L., {Baring}, M.~G., \& {Wijers}, R.~A.~M.~J.
  2016, \mnras, 461, 877, \dodoi{10.1093/mnras/stw1279}

\bibitem[{{Vigan{\`o}} {et~al.}(2013){Vigan{\`o}}, {Rea}, {Pons}, {Perna},
  {Aguilera}, \& {Miralles}}]{2013MNRAS.434..123V}
{Vigan{\`o}}, D., {Rea}, N., {Pons}, J.~A., {et~al.} 2013, \mnras, 434, 123,
  \dodoi{10.1093/mnras/stt1008}

\bibitem[{{Wadiasingh} \& {Timokhin}(2019)}]{2019ApJ...879....4W}
{Wadiasingh}, Z., \& {Timokhin}, A. 2019, \apj, 879, 4,
  \dodoi{10.3847/1538-4357/ab2240}

\bibitem[{{Wang} {et~al.}(2010){Wang}, {Lai}, \& {Han}}]{2010MNRAS.403..569W}
{Wang}, C., {Lai}, D., \& {Han}, J. 2010, \mnras, 403, 569,
  \dodoi{10.1111/j.1365-2966.2009.16074.x}

\bibitem[{{Wang} {et~al.}(2018){Wang}, {Luo}, {Yue}, {Chen}, {Lee}, \&
  {Xu}}]{2018ApJ...852..140W}
{Wang}, W., {Luo}, R., {Yue}, H., {et~al.} 2018, \apj, 852, 140,
  \dodoi{10.3847/1538-4357/aaa025}

\bibitem[{{Wang} {et~al.}(2019){Wang}, {Zhang}, {Chen}, \&
  {Xu}}]{2019ApJ...876L..15W}
{Wang}, W., {Zhang}, B., {Chen}, X., \& {Xu}, R. 2019, \apjl, 876, L15,
  \dodoi{10.3847/2041-8213/ab1aab}

\bibitem[{{Woods} {et~al.}(1999){Woods}, {Kouveliotou}, {van Paradijs},
  {Hurley}, {Kippen}, {Finger}, {Briggs}, {Dieters}, \&
  {Fishman}}]{1999ApJ...519L.139W}
{Woods}, P.~M., {Kouveliotou}, C., {van Paradijs}, J., {et~al.} 1999, \apjl,
  519, L139, \dodoi{10.1086/312124}

\bibitem[{{Yang} \& {Zhang}(2018)}]{2018ApJ...868...31Y}
{Yang}, Y.-P., \& {Zhang}, B. 2018, \apj, 868, 31,
  \dodoi{10.3847/1538-4357/aae685}

\bibitem[{{Younes} {et~al.}(2014){Younes}, {Kouveliotou}, {van der Horst},
  {Baring}, {Granot}, {Watts}, {Bhat}, {Collazzi}, {Gehrels}, {Gorgone},
  {G{\"o}{\u g}{\"u}{\c s}}, {Gruber}, {Grunblatt}, {Huppenkothen}, {Kaneko},
  {von Kienlin}, {van der Klis}, {Lin}, {Mcenery}, {van Putten}, \&
  {Wijers}}]{2014ApJ...785...52Y}
{Younes}, G., {Kouveliotou}, C., {van der Horst}, A.~J., {et~al.} 2014, \apj,
  785, 52, \dodoi{10.1088/0004-637X/785/1/52}

\bibitem[{{Zhang} {et~al.}(2018){Zhang}, {Gajjar}, {Foster}, {Siemion},
  {Cordes}, {Law}, \& {Wang}}]{2018ApJ...866..149Z}
{Zhang}, Y.~G., {Gajjar}, V., {Foster}, G., {et~al.} 2018, \apj, 866, 149,
  \dodoi{10.3847/1538-4357/aadf31}

\end{thebibliography}

\end{document}